\documentclass[aps,prl,reprint,showpacs,superscriptaddress]{revtex4-2}

\usepackage{IEEEtrantools}
\usepackage{amsmath}
\usepackage{amsfonts}
\usepackage{amssymb}
\usepackage{wasysym}
\usepackage[
colorlinks,
linkcolor=blue,
urlcolor=blue,
citecolor=blue,
]{hyperref}
\usepackage{graphics,psfrag}
\usepackage{graphicx,psfrag}
\usepackage{xcolor}
\usepackage{braket}
\usepackage{xspace}
\usepackage{dcolumn}
\usepackage{textcomp}

\newcommand{\etal}{\textit{et al.~}}

\begin{document}
	
	\title{Fast hydrogen atom diffraction through monocrystalline graphene}
	
	\author{Pierre Guichard}\thanks{These authors contributed equally to this work.}
	\affiliation{Université de Strasbourg, CNRS, Institut de Physique et Chimie des Matériaux de Strasbourg, UMR 7504, 67000 Strasbourg, France}

	\author{Arnaud Dochain}\thanks{These authors contributed equally to this work.}
	\affiliation{Institute of Condensed Matter and Nanosciences, Universit\'e
	Catholique de Louvain, B-1348 Louvain-la-Neuve, Belgium}
	
	\author{Rapha\"el Marion}\thanks{These authors contributed equally to this work.}
	\affiliation{Institute of Condensed Matter and Nanosciences, Universit\'e
	Catholique de Louvain, B-1348 Louvain-la-Neuve, Belgium}
	\affiliation{Royal Observatory of Belgium (ROB-ORB), B-1180 Brussels, Belgium}
	
\author{Pauline de Crombrugghe de Picquendaele}
\affiliation{Institute of Condensed Matter and Nanosciences, Universit\'e
	Catholique de Louvain, B-1348 Louvain-la-Neuve, Belgium}	

\author{Nicolas Lejeune}
\affiliation{Institute of Condensed Matter and Nanosciences, Universit\'e
	Catholique de Louvain, B-1348 Louvain-la-Neuve, Belgium}

\author{Beno\^it Hackens}
	\affiliation{Institute of Condensed Matter and Nanosciences, Universit\'e
	Catholique de Louvain, B-1348 Louvain-la-Neuve, Belgium}	
\author{Paul-Antoine Hervieux}\email[]{paul-antoine.hervieux@ipcms.unistra.fr}
\affiliation{Université de Strasbourg, CNRS, Institut de Physique et Chimie des Matériaux de Strasbourg, UMR 7504, 67000 Strasbourg, France}
	
	\author{Xavier Urbain}\email[]{xavier.urbain@uclouvain.be}
	\affiliation{Institute of Condensed Matter and Nanosciences, Universit\'e
		Catholique de Louvain, B-1348 Louvain-la-Neuve, Belgium}

\date{\today}

\begin{abstract}
We report fast atom diffraction through single-layer graphene using hydrogen atoms at kinetic energies from 150 to 1200 eV. High-resolution images reveal overlapping hexagonal patterns from coexisting monocrystalline domains. Time-of-flight tagging confirms negligible energy loss, making the method suitable for matter-wave interferometry. The diffraction is well described by the eikonal approximation, with accurate modeling requiring the full 3D interaction potential from density functional theory. Simpler models fail to reproduce the data, highlighting the exceptional sensitivity of diffraction patterns to atom–surface interactions and their potential for spectroscopic applications.
\end{abstract}

\maketitle

The matter-wave hypothesis enunciated by Louis de Broglie in 1923 \cite{deBroglie1923} rapidly found its experimental verification in the pioneering electron scattering experiments of Thomson and Reid with thin platinum films \cite{Thomson1927}, and Davisson and Germer with mono-crystalline nickel \cite{Davisson1927}, both published in 1927. Those experiments were interpreted as resulting from diffraction of matter waves associated with the electrons, in complete similarity with the patterns recorded earlier with X-rays. Fast electron diffraction has since become the workhorse of many solid state physics laboratories. Over the years, diffraction was observed with all kinds of elementary and composite particles, from neutrons to atomic clusters, both in reflection and transmission through thin films and, more recently, through nanostructured graphene sheets \cite{Brand2015} and laser-light gratings \cite{Nairz2001}, as developed by Arndt and coworkers in the pursuit of interferometry with ever bigger, more complex objects. Atom interferometry also finds application in fundamental physics experiments testing CPT symmetry and the weak equivalence principle \cite{Mueller2020}.

Composite particles pose the extra challenge that their diffuse electron cloud will interact with the equally diffuse electron density permeating the space between ionic centers. This spatial overlap will result in electron exchange, inelastic interactions including charge transfer, and atomic displacement. Such an experiment was performed by Schmidt \etal \cite{Schmidt2008} with fast molecular hydrogen ions colliding with helium atoms, producing Young's slit interference patterns in the momentum imparted to the recoiling helium ion. With the availability of two-dimensional (2D) materials such as graphene, atom diffraction experiments may now be realized in a more straightforward manner with fast atoms impinging on a stationary target, as proposed by Brand \etal \cite{Brand2019} and just realized by Kanitz \etal \cite{Kanitz2025}. We should also mention the calculations of Labaigt \etal \cite{Labaigt2014} who suggested to attempt electron capture imaging of two-dimensional materials by passing fast protons through graphene.

In the present paper, we demonstrate that single crystalline domains may be probed with fast hydrogen atoms transmitted through suspended graphene sheets, as revealed by the direct observation of hexagonal diffraction patterns repeating themselves to high diffraction order. The intensities associated with those successive diffraction orders is confronted to models of the atom-surface interaction potential. To support our findings, we present full-scale DFT calculations, rigorously benchmarked against established methods and approximations.

{\it Experiment} -- The experimental set-up (detailed in Supplemental Material \cite{supplemental}, Sec.~I, Fig.~S1) comprises a duoplasmatron ion source fed with hydrogen gas, an accelerating and focusing column, a Wien filter, a 45~degree cylindrical deflector and a vertical steerer. The latter is used to chop the ion beam for time-of-flight (TOF) spectroscopy. Three circular apertures distributed along the 3~meter long flight path define the emittance down to 0.5~mm$\cdot$mrad. A short gas cell located behind the first aperture and fed with carbon dioxide converts a small fraction of the proton beam into ground state hydrogen atoms. The third aperture 1~mm in diameter is placed right in front of the Cu transmission electron microscopy (TEM) grid supporting the temperature regulated graphene sample (see Supplemental Material \cite{supplemental}, Sec. II, which includes refs. \cite{Tian2016,Meyer2007,Bao2009}). The interaction chamber is evacuated down to $\simeq 2 \times 10^{-9}$ mbar. Scattered particles are detected 93~cm downstream by means of a triple microchannel plate stack 40~mm in diameter backed with a resistive anode. Atoms are counted one at a time with position and arrival time resolution of 50~{\textmu}m and 100~ps, respectively. Typical count rates are 50~Hz to 5000~Hz depending on the accelerating potential ranging from 150~V to 1200~V. The low count rates are due in part to the low detection efficiency of slow atoms, besides the space charge limit and restricted emittance imposed by the apertures.

Different commercial and home-grown suspended single layer graphene samples have been tested with varying production methods. All come with heavy contamination resulting in zero initial transmission except for a central narrow spot corresponding to ballistic trajectories through holes present in the graphene monolayer. Various methods have been devised to actively remove impurities, e.g. laser desorption \cite{Niggas2020} or plasma exposure \cite{Ferrah2016}. We limited ourselves to thermal desorption at moderate temperature, as suggested by hydrogenated graphene studies \cite{Whitener2014}. After 24 hours at 260 $^\circ$C, a diffuse background with a Gaussian like radial distribution appears that is centered around the transmission peak.  This suggests that the diffuse background, besides inelastic collisions with pristine graphene, is caused by adsorbed impurities such as water molecules and/or fabrication residues such as poly(methyl methacrylate) (PMMA). After 48 hours of thermal desorption, a clear diffraction pattern emerges, that gains in contrast over time. Cooling the sample below 100 $^\circ$C invariably results in a reduction of contrast (see Supplemental Material \cite{supplemental}, Sec.~II).

\begin{figure}[h!]
	\includegraphics[width=\linewidth]{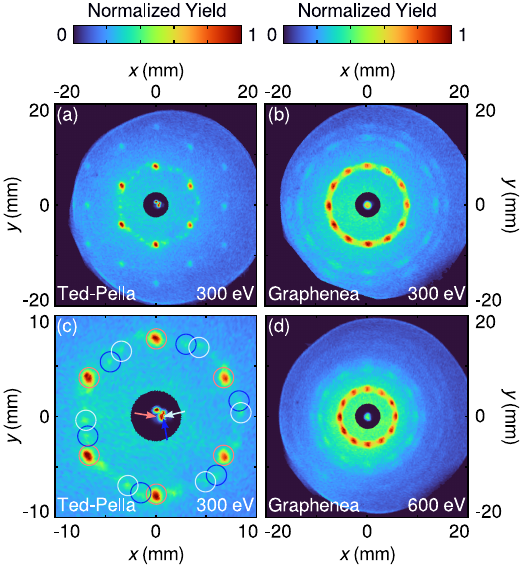}
	\caption{Diffraction images recorded with H atoms impinging on single layer graphene deposited on TEM grids. All images are normalized to the first-order diffraction peak. The yield in the image center (black circle) has been further divided by 1000 to reveal the spatial distribution of ballistic trajectories through holes present in the graphene monolayer.
	(a) 300 eV ({\it Ted Pella}); (b) 300 eV ({\it Graphenea}) ; (c) determination of position and orientation of single crystalline domains ({\it Ted Pella}) -- arrows point to the reconstructed location of the diffracting domains; (d) 600 eV ({\it Graphenea}). }
	\label{fig1}
\end{figure}

{\it Results} -- Several graphene samples (single layer graphene suspended on ultra-fine mesh copper TEM grids with 6.5~{\textmu}m circular holes and a pitch of 12.5~{\textmu}m, {\it Ted Pella}) produced an hexagonal diffraction pattern dominated by six bright spots located at the expected distance from the central feature. As seen in Fig.~\ref{fig1}(a), their position matches the lattice parameters of hexagonal graphene, and scales according to the de Broglie wavelength of the impinging atoms, causing the image size to drop by a factor of two between 150~eV and 600~eV (see Supplemental material \cite{supplemental}, Sec.~III, Fig.~S4). Additional spots are only visible up to the fifth order of diffraction, due either to the limited field of view at low energy, or to the presence of a strong diffuse background at high energy, and to the rapid drop in intensity of successive diffraction orders in all cases.

Secondary hexagonal diffraction patterns appear rotated by some random angle with respect to the dominant orientation (images have been rotated to have the corresponding hexagon pointing upward in the figure).
Reconstructing their center of symmetry (Fig.~\ref{fig1}(c)) reveals that they originate from different parts of the graphene target, suggesting that one could illuminate a single crystalline domain with tighter beam collimation. The composite diffraction image could be used for orientation mapping as performed by low-energy electron diffraction \cite{Zhao2017} and scanning electron microscopy \cite{Neubeck2010,Caplins2019}.

Another class of single layer graphene samples (chemical vapor deposition graphene transferred on Quantifoil\textregistered\ TEM grid with 2~{\textmu}m circular holes and a pitch of 4~{\textmu}m, {\it Graphenea}) produces the typical ring structure of polycrystalline material diffraction images, with the additional appearance of twelve bright spots (Figs. \ref{fig1}(b) and (d)). We interpret this as a result of two dominant orientations locked at $\sim 30$ degree with respect to one another due to better lattice compatibility at grain boundaries \cite{Huang2011}.

The reason for higher orientation disorder observed with {\it Graphenea} targets may be the more complete outgassing of those samples. Indeed, close examination of the central spot of diffraction images recorded with {\it Ted Pella} targets shows stronger inhomogeneity, possibly reflecting a severely reduced pristine graphene area. The multiple bright spots (dark red in Fig.~\ref{fig1}(a)) in the center of the image where zeroth-order diffraction is expected (count rate divided by 1000) correspond to holes through which hydrogen atoms fly unaffected, to be contrasted with the more homogeneous, less intense, central spot visible in Fig.~\ref{fig1}(b).

By rapidly chopping the proton beam, one may simultaneously record the position and time of arrival of individual atoms. We exploit the presence of the ballistic peak in the middle of the image (see Fig.~\ref{fig2}(a)) to calibrate our velocity scale (see Supplemental Material \cite{supplemental}, Sec. I). From the width of the TOF distribution, we infer an energy spread $<1$ eV FWHM at 150 eV. Interestingly, the diffuse scattering background comes with a significant energy loss (14$\pm$1 eV at 600 eV), while the diffraction spots are not distinguishable, within statistics, from the ballistic peak in terms of TOF (Fig.~\ref{fig2}(b)), their energy loss not exceeding 1~eV at 1200 eV impact energy (see Supplemental Material \cite{supplemental}, Sec. III). Filtering the raw image on the basis of energy loss produces distinct patterns corresponding to inelastic scattering (Filter 1, Fig.~\ref{fig2}(c)) and diffraction (Filter 2, Fig.~\ref{fig2}(d)).

\begin{figure}[h!]
	\includegraphics[width=\columnwidth]{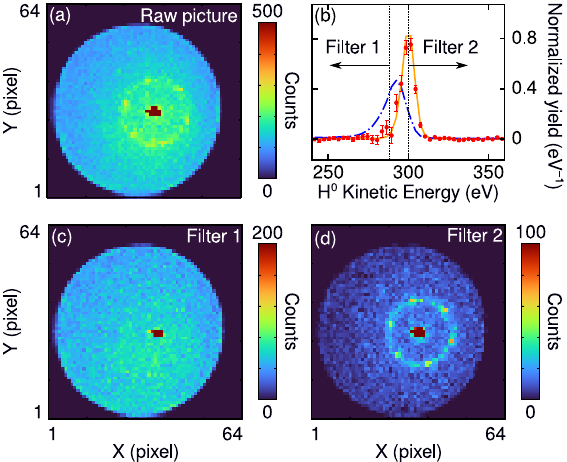}
	\caption{Sorting of elastic and inelastic scattering events. (a) Raw picture (300 eV, Ted Pella); (b) Kinetic energy distribution of H atoms transmitted through graphene -- full line: ballistic peak, dot-dashed line: diffuse background, symbols: diffraction peaks; (c) Inelastic scattering events ($E\leq 288$ eV); (d) Elastic scattering events ($E\geq 300$ eV). The diffraction signal in (b) was obtained by subtracting the kinetic energy distribution recorded in an adjacent area of similar size. No subtraction is needed to obtain the filtered images. All images in these panels were binned to $64\times 64$ pixels to improve statistics.}
	\label{fig2}
\end{figure}

The high temperature at which diffraction images were acquired raises the issue of loss of contrast due to lattice thermal motion. This effect is usually taken into account with the so-called Debye-Waller factor \cite{Shevitski2013} (see Supplemental Material \cite{supplemental}, Sec.~IV). We experimentally investigated the sensitivity of the diffraction pattern to the temperature of the sample, which we varied between 100~$^\circ$C and 260~$^\circ$C. No significant change could be observed over that temperature range in the intensity ratios recorded at 300 eV for the first three orders of diffraction. Higher diffraction orders and an extended temperature range are obviously needed to quantify the effect of lattice vibrations.

{\it Theory} -- In order to evaluate the diffraction pattern produced by the coherent scattering of hydrogen atoms through the graphene sheet, the eikonal approximation is used here. The eikonal approximation allows us to reduce the scattering problem to the modeling of the interaction potential $V$ integrated along $z$. The range of validity of this approximation \cite{Landau1981} makes it suitable for the present study. Inelastic processes are assumed to be weak. Indeed, according to the study by Ehemann \etal \cite{Ehemann2012}, for hydrogen atoms impacting a graphene sheet at normal incidence with kinetic energies greater than $200$ eV, the transmission through the surface should be nearly complete.

\begin{figure}[!ht]
	\centering
	\includegraphics[width=\columnwidth]{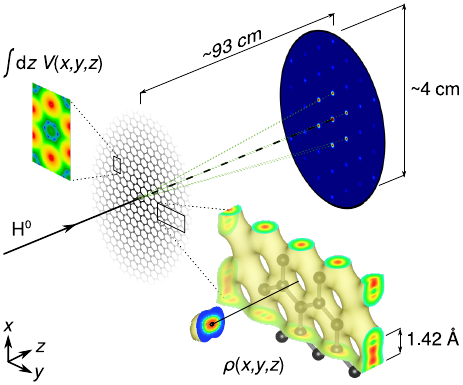}
	\caption{Schematic view of the modeling of coherent diffraction of a beam of hydrogen atoms through a sheet of graphene (see text).}
	\label{fig:theory-schematic}
\end{figure}

 As shown in Fig.~\ref{fig:theory-schematic}, the atomic beam propagates along the $z$ axis and the graphene sheet lies in the $xy$ plane. Using the symmetry properties of graphene, the intensity of the atomic beam is given by \cite{Newton1982}:

\begin{align}
	I_{m, n}&= \nonumber\\
	&\left|\frac{1}{\mathcal{A}} \iint_{\Omega_x, \Omega_y} \mathrm{d} x \mathrm{~d} y ~\text{e}^{-i G_x x-i G_y y} ~\text{e}^{-\frac{i}{\hbar v_z} \int_{-\infty}^{+\infty} \mathrm{d} z V(x, y, z)}\right|^2,
\end{align}

where $\mathcal{A} \equiv L_x L_y$ is the area of the unit cell and $\Omega_{x,y}$ are the domains of integration in $x$ and $y$. The reciprocal lattice vectors are $G_x = 2\pi m/L_x$ and $G_y = 2\pi n/L_y$ and the incoming velocity along the $z$ axis is $v_z$. Finally, the potential that describes the interaction between a hydrogen atom at position $(x,y,z)$ and the graphene sheet is $V(x,y,z)$.

The interaction of the hydrogen atom with the graphene sheet can be described theoretically with varying levels of approximation. One sophisticated way is the Density Functional Theory (DFT). In this model, for a given position of the hydrogen atom with respect to the surface, the electron density $\rho (x,y,z)$ of the whole combination of graphene and atom is calculated. The electronic structure of the latter is modified by the presence of the surface and vice versa. Such calculation is non-perturbative and uses the Born-Oppenheimer approximation.

In what follows, this global {\it ab initio} treatment will be our reference model and we will show that only a theoretical description at this level of accuracy brings it in satisfactory agreement with the experimental data. To demonstrate this, we will compare the diffraction patterns obtained using this approach with those obtained using other robust but less accurate approaches.

The present {\it ab initio} calculations were performed using the {\sc Quantum ESPRESSO} software suite \cite{QE-2009, QE-2017, QE-2020}. Pseudopotentials based on the Perdew-Burke-Ernzerhof (PBE) exchange correlation functional were employed to describe the carbon and hydrogen atoms (see Supplemental Material \cite{supplemental}, Sec.~V). The total energy of the system is calculated as a function of the position of the hydrogen atom relative to the graphene surface. It includes contributions from the electronic structure of the graphene sheet, the hydrogen atom and their mutual interaction. To calculate the latter, the energies of the isolated graphene sheet and the hydrogen atom are subtracted from the total energy of the system.

In order to demonstrate that an accurate description of the atom-surface interaction is required, we have used an interaction potential constructed from the H-C binary potential. In this model, the total H-graphene interaction potential is the sum of the binary H-C potential resulting from all the carbon atoms in the graphene sheet and is written as
\begin{equation}
	V(x,y,z)=\sum_{i} V_{\mathrm{H-C}}(x-x_i,y-y_i,z-z_i) \;.
\end{equation}
$V_{\mathrm{H-C}}$ is calculated using DFT and represents the interaction energy of the hydrogen atom at a given distance relative to the $i$-th carbon atom. We have checked that the values of the equilibrium position and the depth of the CH potential well are compatible with those found in the literature \cite{Vermeeren2022}.

\begin{figure}[!ht]
	\centering
	\includegraphics[width=\columnwidth]{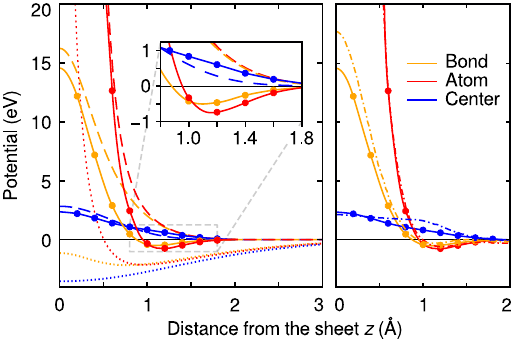}
	\caption{Hydrogen-graphene interaction potential as a function of $z$ ($z=0$ corresponds to the surface) for three positions in the $xy$ plane: at the center of the carbon ring (in blue, {\it Center}), facing a carbon atom (in red, {\it Atom}) and in the middle of a carbon-carbon bond (in orange, {\it Bond}). The different models used to generate these potentials are {\it ab initio} (solid lines), Brand \etal \cite{Brand2019} (dashed lines), H-C binary (divided by~10, dotted lines) and SCC-DFTB \cite{Ehemann2012} (dash dotted lines).}
	\label{fig:PEC}
\end{figure}

To date, there are two significant studies, by Ehemann \etal \cite{Ehemann2012} and Brand \etal \cite{Brand2019}, who have studied the interaction potential between a hydrogen atom and a graphene surface, each using different computational methods. Brand \etal \cite{Brand2019} used a hybrid approach to calculate the 3D interaction potential. Their method starts with a full DFT calculation for a few selected points on the lattice. They then used this information to approximate the potential at any other point on the lattice by fitting a proportionality constant to the graphene electron density computed in the absence of hydrogen atom. The trade-off, however, is that the extrapolated potential, which is not directly derived from a full DFT calculation, can lead to some imprecision, especially in the regions that are not well sampled by the initial points calculated by DFT. Ehemann \etal \cite{Ehemann2012} used a different approach based on the Self-Consistent-Charge Density-Functional Tight-Binding (SCC-DFTB) method \cite{Elstner1998}. It provides an efficient and scalable method for simulating the electronic structure of large systems. Compared to full DFT methods, SCC-DFTB is less accurate but much faster. Ehemann \etal provide information only on the potential computed at three specific positions on the graphene lattice ({\it Bond}, {\it Atom} and {\it Center}) as a function of $z$, the distance from the surface. However, in order to determine the diffraction pattern one needs to have access to the potential calculated at all lattice points, hence we could not generate the corresponding diffraction images.

Figure~\ref{fig:PEC} shows the interaction potential as a function of the distance $z$ from the hydrogen atom to the graphene surface, as computed with the different methods. At the center of the carbon ring, the Brand \etal potential is very close to the {\it ab initio} potential, but significant differences are observed at the middle of the C-C bonds or facing the carbon atoms. At the latter position, the {\it ab initio} potential shows negative values around $z=1$~{\AA}, allowing hydrogen atoms to be adsorbed onto the graphene surface, in agreement with the predictions of SCC-DFTB \cite{Ehemann2012} and the DFT calculations of Ivanovskaya \etal \cite{Ivanovskaya2010}.

\begin{figure}[!h]
	\centering
	\includegraphics[width=\columnwidth]{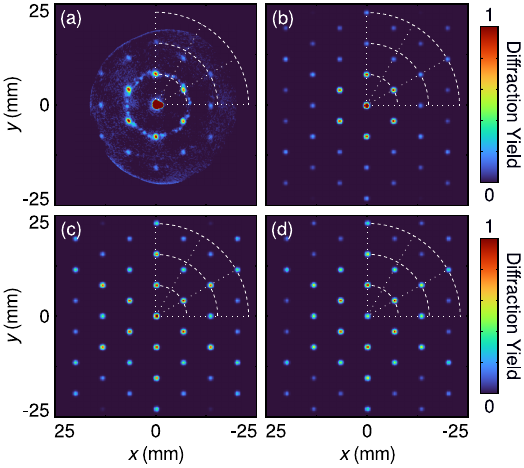}
	\caption{Diffraction pattern of a beam of hydrogen atoms of kinetic energy $E=300$~eV through a graphene sheet. Theoretical predictions obtained using the eikonal approximation and three interaction potential models are compared with experimental measurements: (a) experiment (diffusion background subtracted), (b) {\it ab initio}, (c) H-C binary, and (d) Brand \etal \cite{Brand2019}.}
	\label{fig:diffraction-map-theo-exp-compare}
\end{figure}

Fig.~\ref{fig:diffraction-map-theo-exp-compare} and Table~ \ref{tbl:theo-exp-diffraction-order} compare the experimental and simulated diffraction patterns and relative peak intensities. Both data show very clearly that the best agreement between theory and experiment is obtained when using the {\it ab initio} potential calculated in all space. The approximation to the interaction potential made by Brand \etal \cite{Brand2019} based on the total electron density does not give good results, nor does the approximation based on isolated atom-atom interactions. The former neglects both the modification of C-C chemical bonds due to the presence of the hydrogen atom and the modification of hydrogen atom orbitals due to the presence of graphene, while the latter vastly overestimates the atom-lattice interaction by ignoring the alteration of the C-H potential due to the C-C bonds. The introduction of the Debye-Waller factor does not modify this conclusion (see Supplemental Material \cite{supplemental}, Sec. IV, Table SIII). This shows how important it is to treat the H-graphene interaction microscopically and across the whole space in order to describe the diffraction pattern.

\begin{table}[!ht]
	\begin{tabular*}{\columnwidth}{l @{\extracolsep{\fill}} c @{\extracolsep{\fill}} c @{\extracolsep{\fill}} c @{\extracolsep{\fill}} c}
		\hline\hline
		\multicolumn{5}{c}{$300~$eV}\\
		$G$ & Experiment & {\it ab initio} & H-C binary & Brand \etal \cite{Brand2019}\\
		\hline
		$1$        & 1 & 1 & 1 & 1 \\
		$\sqrt{3}$ & 0.241(20) & 0.159 & 0.455 & 0.529 \\
		$2$        & 0.160(19) & 0.200 & 0.848 & 0.621 \\
		\hline\hline
		\multicolumn{5}{c}{$600~$eV}\\
		$G$ & Experiment & {\it ab initio} & H-C binary & Brand \etal \cite{Brand2019}\\
		\hline
		$1$        & 1 & 1 & 1 & 1 \\
		$\sqrt{3}$ & 0.258(32) & 0.166 & 0.412 & 0.536 \\
		$2$        & 0.168(29) & 0.155 & 0.423 & 0.257 \\
		$\sqrt{7}$ & 0.164(26) & 0.095 & 0.149 & 0.154 \\
		$3$        & 0.114(27) & 0.186 & 0.290 & 0.253 \\

		\hline\hline
	\end{tabular*}
	
	\caption{Relative diffraction intensities (normalized to first order) at 300 and 600~eV as observed and predicted with different theoretical models. $G$ is given in units of $G_{\text min}=4\pi/\sqrt{3}a$, with $a=246$~pm the lattice parameter. Numbers in parentheses represent 1$\sigma$ uncertainty in units of the last significant digits.}
	
	\label{tbl:theo-exp-diffraction-order}
	
\end{table}

{\it Conclusions} -- Experiments presented here demonstrate the coherent diffraction of fast hydrogen atoms through free-standing mono- and polycrystalline graphene samples. Time-of-flight measurements confirm the elastic character of diffraction events, while inelastic scattering events are characterized by sizable energy loss. A varying degree of surface contamination may explain why a single orientation dominates the recorded patterns in some cases, while the whole orientation spectrum is adding up to create circular images with distinct accumulation islands for other graphene samples.

Compared to ionic projectiles such as protons, neutral atoms have the advantage of avoiding the excitation of plasmonic resonances which inevitably dominate the interaction of charged particles with the graphene electron density.
Neutral projectiles open up additional possibilities, among them the study of insulating 2D materials such as hexagonal boron nitride (hBN).

In contrast to electron diffraction, three-dimensional (3D) interaction potential calculations demonstrate that the atom–surface interaction potential is determined not solely by the electron density, but also by pronounced polarization effects originating from the mutual perturbation between the hydrogen electron cloud and the graphene sheet. The remarkable sensitivity of the relative intensities of successive diffraction orders to subtle features of this potential opens the door to a novel form of atom-surface interaction spectroscopy.

{\it Acknowledgments} -- Experiments have been funded by the Fonds de la Recherche Scientifique - FNRS under contract No. 4.4504.10 and the F\'ed\'eration Wallonie-Bruxelles through ARC Grants No. 16/21-077 and No. 21/26-116. This
work was also supported by the FLAG-ERA grant TATTOO. X.U. and B.H. are Senior Research Associates of the Fonds de la Recherche Scientifique - FNRS. A.D. acknowledges support from the Belgian State for the grant allocated by Royal Decree for research in the domain of controlled thermonuclear fusion. The theoretical work was funded by the French National Research Agency (ANR) through the Programme d'Investissement d'Avenir under contract ANR-11-LABX-0058\_NIE and ANR-17-EURE-0024 within the Investissement d'Avenir program ANR-10-IDEX-0002-02. The authors would like to acknowledge the High Performance Computing Center of the University of Strasbourg for supporting this work by providing scientific support and access to computing resources. Part of the computing resources were funded by the Equipex Equip@Meso project (Programme Investissements d'Avenir) and the CPER Alsacalcul/Big Data. P. G. and P.-A. H. would like to thank R{\'e}mi Pasquier for his advice on the use of the DFT code.

\bibliography{bibliography}

\begin{thebibliography}{5}%
\makeatletter
\providecommand \@ifxundefined [1]{%
 \@ifx{#1\undefined}
}%
\providecommand \@ifnum [1]{%
 \ifnum #1\expandafter \@firstoftwo
 \else \expandafter \@secondoftwo
 \fi
}%
\providecommand \@ifx [1]{%
 \ifx #1\expandafter \@firstoftwo
 \else \expandafter \@secondoftwo
 \fi
}%
\providecommand \natexlab [1]{#1}%
\providecommand \enquote  [1]{``#1''}%
\providecommand \bibnamefont  [1]{#1}%
\providecommand \bibfnamefont [1]{#1}%
\providecommand \citenamefont [1]{#1}%
\providecommand \href@noop [0]{\@secondoftwo}%
\providecommand \href [0]{\begingroup \@sanitize@url \@href}%
\providecommand \@href[1]{\@@startlink{#1}\@@href}%
\providecommand \@@href[1]{\endgroup#1\@@endlink}%
\providecommand \@sanitize@url [0]{\catcode `\\12\catcode `\$12\catcode
  `\&12\catcode `\#12\catcode `\^12\catcode `\_12\catcode `\%12\relax}%
\providecommand \@@startlink[1]{}%
\providecommand \@@endlink[0]{}%
\providecommand \url  [0]{\begingroup\@sanitize@url \@url }%
\providecommand \@url [1]{\endgroup\@href {#1}{\urlprefix }}%
\providecommand \urlprefix  [0]{URL }%
\providecommand \Eprint [0]{\href }%
\providecommand \doibase [0]{https://doi.org/}%
\providecommand \selectlanguage [0]{\@gobble}%
\providecommand \bibinfo  [0]{\@secondoftwo}%
\providecommand \bibfield  [0]{\@secondoftwo}%
\providecommand \translation [1]{[#1]}%
\providecommand \BibitemOpen [0]{}%
\providecommand \bibitemStop [0]{}%
\providecommand \bibitemNoStop [0]{.\EOS\space}%
\providecommand \EOS [0]{\spacefactor3000\relax}%
\providecommand \BibitemShut  [1]{\csname bibitem#1\endcsname}%
\let\auto@bib@innerbib\@empty
\bibitem [{\citenamefont {Tian}\ \emph {et~al.}(2016)\citenamefont {Tian},
  \citenamefont {Yang}, \citenamefont {Liu}, \citenamefont {Wang},
  \citenamefont {Pan}, \citenamefont {Gu},\ and\ \citenamefont
  {Li}}]{Tian2016}%
  \BibitemOpen
  \bibfield  {author} {\bibinfo {author} {\bibfnamefont {S.}~\bibnamefont
  {Tian}}, \bibinfo {author} {\bibfnamefont {Y.}~\bibnamefont {Yang}}, \bibinfo
  {author} {\bibfnamefont {Z.}~\bibnamefont {Liu}}, \bibinfo {author}
  {\bibfnamefont {C.}~\bibnamefont {Wang}}, \bibinfo {author} {\bibfnamefont
  {R.}~\bibnamefont {Pan}}, \bibinfo {author} {\bibfnamefont {C.}~\bibnamefont
  {Gu}},\ and\ \bibinfo {author} {\bibfnamefont {J.}~\bibnamefont {Li}},\
  }\bibfield  {title} {\bibinfo {title} {Temperature-dependent {Raman}
  investigation on suspended graphene: Contribution from thermal expansion
  coefficient mismatch between graphene and substrate},\ }\href
  {https://doi.org/10.1016/j.carbon.2016.03.046} {\bibfield  {journal}
  {\bibinfo  {journal} {Carbon}\ }\textbf {\bibinfo {volume} {104}},\ \bibinfo
  {pages} {27–32} (\bibinfo {year} {2016})}\BibitemShut {NoStop}%
\bibitem [{\citenamefont {Meyer}\ \emph {et~al.}(2007)\citenamefont {Meyer},
  \citenamefont {Geim}, \citenamefont {Katsnelson}, \citenamefont {Novoselov},
  \citenamefont {Booth},\ and\ \citenamefont {Roth}}]{Meyer2007}%
  \BibitemOpen
  \bibfield  {author} {\bibinfo {author} {\bibfnamefont {J.~C.}\ \bibnamefont
  {Meyer}}, \bibinfo {author} {\bibfnamefont {A.~K.}\ \bibnamefont {Geim}},
  \bibinfo {author} {\bibfnamefont {M.~I.}\ \bibnamefont {Katsnelson}},
  \bibinfo {author} {\bibfnamefont {K.~S.}\ \bibnamefont {Novoselov}}, \bibinfo
  {author} {\bibfnamefont {T.~J.}\ \bibnamefont {Booth}},\ and\ \bibinfo
  {author} {\bibfnamefont {S.}~\bibnamefont {Roth}},\ }\bibfield  {title}
  {\bibinfo {title} {The structure of suspended graphene sheets},\ }\href
  {https://doi.org/10.1038/nature05545} {\bibfield  {journal} {\bibinfo
  {journal} {Nature}\ }\textbf {\bibinfo {volume} {446}},\ \bibinfo {pages}
  {60–63} (\bibinfo {year} {2007})}\BibitemShut {NoStop}%
\bibitem [{\citenamefont {Bao}\ \emph {et~al.}(2009)\citenamefont {Bao},
  \citenamefont {Miao}, \citenamefont {Chen}, \citenamefont {Zhang},
  \citenamefont {Jang}, \citenamefont {Dames},\ and\ \citenamefont
  {Lau}}]{Bao2009}%
  \BibitemOpen
  \bibfield  {author} {\bibinfo {author} {\bibfnamefont {W.}~\bibnamefont
  {Bao}}, \bibinfo {author} {\bibfnamefont {F.}~\bibnamefont {Miao}}, \bibinfo
  {author} {\bibfnamefont {Z.}~\bibnamefont {Chen}}, \bibinfo {author}
  {\bibfnamefont {H.}~\bibnamefont {Zhang}}, \bibinfo {author} {\bibfnamefont
  {W.}~\bibnamefont {Jang}}, \bibinfo {author} {\bibfnamefont {C.}~\bibnamefont
  {Dames}},\ and\ \bibinfo {author} {\bibfnamefont {C.~N.}\ \bibnamefont
  {Lau}},\ }\bibfield  {title} {\bibinfo {title} {Controlled ripple texturing
  of suspended graphene and ultrathin graphite membranes},\ }\href
  {https://doi.org/10.1038/nnano.2009.191} {\bibfield  {journal} {\bibinfo
  {journal} {Nature Nanotechnology}\ }\textbf {\bibinfo {volume} {4}},\
  \bibinfo {pages} {562–566} (\bibinfo {year} {2009})}\BibitemShut {NoStop}%
\bibitem [{\citenamefont {Shevitski}\ \emph {et~al.}(2013)\citenamefont
  {Shevitski}, \citenamefont {Mecklenburg}, \citenamefont {Hubbard},
  \citenamefont {White}, \citenamefont {Dawson}, \citenamefont {Lodge},
  \citenamefont {Ishigami},\ and\ \citenamefont {Regan}}]{Shevitski2013}%
  \BibitemOpen
  \bibfield  {author} {\bibinfo {author} {\bibfnamefont {B.}~\bibnamefont
  {Shevitski}}, \bibinfo {author} {\bibfnamefont {M.}~\bibnamefont
  {Mecklenburg}}, \bibinfo {author} {\bibfnamefont {W.~A.}\ \bibnamefont
  {Hubbard}}, \bibinfo {author} {\bibfnamefont {E.~R.}\ \bibnamefont {White}},
  \bibinfo {author} {\bibfnamefont {B.}~\bibnamefont {Dawson}}, \bibinfo
  {author} {\bibfnamefont {M.~S.}\ \bibnamefont {Lodge}}, \bibinfo {author}
  {\bibfnamefont {M.}~\bibnamefont {Ishigami}},\ and\ \bibinfo {author}
  {\bibfnamefont {B.~C.}\ \bibnamefont {Regan}},\ }\bibfield  {title} {\bibinfo
  {title} {Dark-field transmission electron microscopy and the {Debye}-{Waller}
  factor of graphene},\ }\href {https://doi.org/10.1103/PhysRevB.87.045417}
  {\bibfield  {journal} {\bibinfo  {journal} {Phys. Rev. B}\ }\textbf {\bibinfo
  {volume} {87}},\ \bibinfo {pages} {045417} (\bibinfo {year}
  {2013})}\BibitemShut {NoStop}%
\bibitem [{\citenamefont {Brand}\ \emph {et~al.}(2019)\citenamefont {Brand},
  \citenamefont {Debiossac}, \citenamefont {Susi}, \citenamefont {Aguillon},
  \citenamefont {Kotakoski}, \citenamefont {Roncin},\ and\ \citenamefont
  {Arndt}}]{Brand2019}%
  \BibitemOpen
  \bibfield  {author} {\bibinfo {author} {\bibfnamefont {C.}~\bibnamefont
  {Brand}}, \bibinfo {author} {\bibfnamefont {M.}~\bibnamefont {Debiossac}},
  \bibinfo {author} {\bibfnamefont {T.}~\bibnamefont {Susi}}, \bibinfo {author}
  {\bibfnamefont {F.}~\bibnamefont {Aguillon}}, \bibinfo {author}
  {\bibfnamefont {J.}~\bibnamefont {Kotakoski}}, \bibinfo {author}
  {\bibfnamefont {P.}~\bibnamefont {Roncin}},\ and\ \bibinfo {author}
  {\bibfnamefont {M.}~\bibnamefont {Arndt}},\ }\bibfield  {title} {\bibinfo
  {title} {Coherent diffraction of hydrogen through the 246 pm lattice of
  graphene},\ }\href {https://doi.org/10.1088/1367-2630/ab05ed} {\bibfield
  {journal} {\bibinfo  {journal} {New Journal of Physics}\ }\textbf {\bibinfo
  {volume} {21}},\ \bibinfo {pages} {033004} (\bibinfo {year}
  {2019})}\BibitemShut {NoStop}%
\end{thebibliography}%


\begin{thebibliography}{31}%
\makeatletter
\providecommand \@ifxundefined [1]{%
 \@ifx{#1\undefined}
}%
\providecommand \@ifnum [1]{%
 \ifnum #1\expandafter \@firstoftwo
 \else \expandafter \@secondoftwo
 \fi
}%
\providecommand \@ifx [1]{%
 \ifx #1\expandafter \@firstoftwo
 \else \expandafter \@secondoftwo
 \fi
}%
\providecommand \natexlab [1]{#1}%
\providecommand \enquote  [1]{``#1''}%
\providecommand \bibnamefont  [1]{#1}%
\providecommand \bibfnamefont [1]{#1}%
\providecommand \citenamefont [1]{#1}%
\providecommand \href@noop [0]{\@secondoftwo}%
\providecommand \href [0]{\begingroup \@sanitize@url \@href}%
\providecommand \@href[1]{\@@startlink{#1}\@@href}%
\providecommand \@@href[1]{\endgroup#1\@@endlink}%
\providecommand \@sanitize@url [0]{\catcode `\\12\catcode `\$12\catcode
  `\&12\catcode `\#12\catcode `\^12\catcode `\_12\catcode `\%12\relax}%
\providecommand \@@startlink[1]{}%
\providecommand \@@endlink[0]{}%
\providecommand \url  [0]{\begingroup\@sanitize@url \@url }%
\providecommand \@url [1]{\endgroup\@href {#1}{\urlprefix }}%
\providecommand \urlprefix  [0]{URL }%
\providecommand \Eprint [0]{\href }%
\providecommand \doibase [0]{https://doi.org/}%
\providecommand \selectlanguage [0]{\@gobble}%
\providecommand \bibinfo  [0]{\@secondoftwo}%
\providecommand \bibfield  [0]{\@secondoftwo}%
\providecommand \translation [1]{[#1]}%
\providecommand \BibitemOpen [0]{}%
\providecommand \bibitemStop [0]{}%
\providecommand \bibitemNoStop [0]{.\EOS\space}%
\providecommand \EOS [0]{\spacefactor3000\relax}%
\providecommand \BibitemShut  [1]{\csname bibitem#1\endcsname}%
\let\auto@bib@innerbib\@empty
\bibitem [{\citenamefont {de~Broglie}(1923)}]{deBroglie1923}%
  \BibitemOpen
  \bibfield  {author} {\bibinfo {author} {\bibfnamefont {L.}~\bibnamefont
  {de~Broglie}},\ }\bibfield  {title} {\bibinfo {title} {{{Waves}} and
  {Quanta}},\ }\href {https://doi.org/10.1038/112540a0} {\bibfield  {journal}
  {\bibinfo  {journal} {Nature}\ }\textbf {\bibinfo {volume} {112}},\ \bibinfo
  {pages} {540} (\bibinfo {year} {1923})}\BibitemShut {NoStop}%
\bibitem [{\citenamefont {Thomson}\ and\ \citenamefont
  {Reid}(1927)}]{Thomson1927}%
  \BibitemOpen
  \bibfield  {author} {\bibinfo {author} {\bibfnamefont {G.~P.}\ \bibnamefont
  {Thomson}}\ and\ \bibinfo {author} {\bibfnamefont {A.}~\bibnamefont {Reid}},\
  }\bibfield  {title} {\bibinfo {title} {{Diffraction of Cathode Rays by a Thin
  Film}},\ }\href {https://doi.org/10.1038/119890a0} {\bibfield  {journal}
  {\bibinfo  {journal} {Nature}\ }\textbf {\bibinfo {volume} {119}},\ \bibinfo
  {pages} {890} (\bibinfo {year} {1927})}\BibitemShut {NoStop}%
\bibitem [{\citenamefont {Davisson}\ and\ \citenamefont
  {Germer}(1927)}]{Davisson1927}%
  \BibitemOpen
  \bibfield  {author} {\bibinfo {author} {\bibfnamefont {C.}~\bibnamefont
  {Davisson}}\ and\ \bibinfo {author} {\bibfnamefont {L.~H.}\ \bibnamefont
  {Germer}},\ }\bibfield  {title} {\bibinfo {title} {{The Scattering of
  Electrons by a Single Crystal of Nickel}},\ }\href
  {https://doi.org/10.1038/119558a0} {\bibfield  {journal} {\bibinfo  {journal}
  {Nature}\ }\textbf {\bibinfo {volume} {119}},\ \bibinfo {pages} {558}
  (\bibinfo {year} {1927})}\BibitemShut {NoStop}%
\bibitem [{\citenamefont {Brand}\ \emph {et~al.}(2015)\citenamefont {Brand},
  \citenamefont {Sclafani}, \citenamefont {Knobloch}, \citenamefont {Lilach},
  \citenamefont {Juffmann}, \citenamefont {Kotakoski}, \citenamefont {Mangler},
  \citenamefont {Winter}, \citenamefont {Turchanin}, \citenamefont {Meyer},
  \citenamefont {Cheshnovsky},\ and\ \citenamefont {Arndt}}]{Brand2015}%
  \BibitemOpen
  \bibfield  {author} {\bibinfo {author} {\bibfnamefont {C.}~\bibnamefont
  {Brand}}, \bibinfo {author} {\bibfnamefont {M.}~\bibnamefont {Sclafani}},
  \bibinfo {author} {\bibfnamefont {C.}~\bibnamefont {Knobloch}}, \bibinfo
  {author} {\bibfnamefont {Y.}~\bibnamefont {Lilach}}, \bibinfo {author}
  {\bibfnamefont {T.}~\bibnamefont {Juffmann}}, \bibinfo {author}
  {\bibfnamefont {J.}~\bibnamefont {Kotakoski}}, \bibinfo {author}
  {\bibfnamefont {C.}~\bibnamefont {Mangler}}, \bibinfo {author} {\bibfnamefont
  {A.}~\bibnamefont {Winter}}, \bibinfo {author} {\bibfnamefont
  {A.}~\bibnamefont {Turchanin}}, \bibinfo {author} {\bibfnamefont
  {J.}~\bibnamefont {Meyer}}, \bibinfo {author} {\bibfnamefont
  {O.}~\bibnamefont {Cheshnovsky}},\ and\ \bibinfo {author} {\bibfnamefont
  {M.}~\bibnamefont {Arndt}},\ }\bibfield  {title} {\bibinfo {title} {An
  atomically thin matter-wave beamsplitter},\ }\href
  {https://doi.org/10.1038/nnano.2015.179} {\bibfield  {journal} {\bibinfo
  {journal} {Nature Nanotechnology}\ }\textbf {\bibinfo {volume} {10}},\
  \bibinfo {pages} {845} (\bibinfo {year} {2015})}\BibitemShut {NoStop}%
\bibitem [{\citenamefont {Nairz}\ \emph {et~al.}(2001)\citenamefont {Nairz},
  \citenamefont {Brezger}, \citenamefont {Arndt},\ and\ \citenamefont
  {Zeilinger}}]{Nairz2001}%
  \BibitemOpen
  \bibfield  {author} {\bibinfo {author} {\bibfnamefont {O.}~\bibnamefont
  {Nairz}}, \bibinfo {author} {\bibfnamefont {B.}~\bibnamefont {Brezger}},
  \bibinfo {author} {\bibfnamefont {M.}~\bibnamefont {Arndt}},\ and\ \bibinfo
  {author} {\bibfnamefont {A.}~\bibnamefont {Zeilinger}},\ }\bibfield  {title}
  {\bibinfo {title} {{Diffraction of Complex Molecules by Structures Made of
  Light}},\ }\href {https://doi.org/10.1103/PhysRevLett.87.160401} {\bibfield
  {journal} {\bibinfo  {journal} {Phys. Rev. Lett.}\ }\textbf {\bibinfo
  {volume} {87}},\ \bibinfo {pages} {160401} (\bibinfo {year}
  {2001})}\BibitemShut {NoStop}%
\bibitem [{\citenamefont {Müller}\ \emph {et~al.}(2020)\citenamefont
  {Müller}, \citenamefont {Nedelec},\ and\ \citenamefont
  {Oberthaler}}]{Mueller2020}%
  \BibitemOpen
  \bibfield  {author} {\bibinfo {author} {\bibfnamefont {S.~R.}\ \bibnamefont
  {Müller}}, \bibinfo {author} {\bibfnamefont {P.}~\bibnamefont {Nedelec}},\
  and\ \bibinfo {author} {\bibfnamefont {M.~K.}\ \bibnamefont {Oberthaler}},\
  }\bibfield  {title} {\bibinfo {title} {From classical xenon fringes to
  hydrogen interferometry},\ }\href {https://doi.org/10.1088/1367-2630/ab9bc1}
  {\bibfield  {journal} {\bibinfo  {journal} {New Journal of Physics}\ }\textbf
  {\bibinfo {volume} {22}},\ \bibinfo {pages} {073060} (\bibinfo {year}
  {2020})}\BibitemShut {NoStop}%
\bibitem [{\citenamefont {Schmidt}\ \emph {et~al.}(2008)\citenamefont
  {Schmidt}, \citenamefont {Sch\"ossler}, \citenamefont {Afaneh}, \citenamefont
  {Sch\"offler}, \citenamefont {Stiebing}, \citenamefont {Schmidt-B\"ocking},\
  and\ \citenamefont {D\"orner}}]{Schmidt2008}%
  \BibitemOpen
  \bibfield  {author} {\bibinfo {author} {\bibfnamefont {L.~P.~H.}\
  \bibnamefont {Schmidt}}, \bibinfo {author} {\bibfnamefont {S.}~\bibnamefont
  {Sch\"ossler}}, \bibinfo {author} {\bibfnamefont {F.}~\bibnamefont {Afaneh}},
  \bibinfo {author} {\bibfnamefont {M.}~\bibnamefont {Sch\"offler}}, \bibinfo
  {author} {\bibfnamefont {K.~E.}\ \bibnamefont {Stiebing}}, \bibinfo {author}
  {\bibfnamefont {H.}~\bibnamefont {Schmidt-B\"ocking}},\ and\ \bibinfo
  {author} {\bibfnamefont {R.}~\bibnamefont {D\"orner}},\ }\bibfield  {title}
  {\bibinfo {title} {{Young-Type Interference in Collisions between Hydrogen
  Molecular Ions and Helium}},\ }\href
  {https://doi.org/10.1103/PhysRevLett.101.173202} {\bibfield  {journal}
  {\bibinfo  {journal} {Phys. Rev. Lett.}\ }\textbf {\bibinfo {volume} {101}},\
  \bibinfo {pages} {173202} (\bibinfo {year} {2008})}\BibitemShut {NoStop}%
\bibitem [{\citenamefont {Brand}\ \emph {et~al.}(2019)\citenamefont {Brand},
  \citenamefont {Debiossac}, \citenamefont {Susi}, \citenamefont {Aguillon},
  \citenamefont {Kotakoski}, \citenamefont {Roncin},\ and\ \citenamefont
  {Arndt}}]{Brand2019}%
  \BibitemOpen
  \bibfield  {author} {\bibinfo {author} {\bibfnamefont {C.}~\bibnamefont
  {Brand}}, \bibinfo {author} {\bibfnamefont {M.}~\bibnamefont {Debiossac}},
  \bibinfo {author} {\bibfnamefont {T.}~\bibnamefont {Susi}}, \bibinfo {author}
  {\bibfnamefont {F.}~\bibnamefont {Aguillon}}, \bibinfo {author}
  {\bibfnamefont {J.}~\bibnamefont {Kotakoski}}, \bibinfo {author}
  {\bibfnamefont {P.}~\bibnamefont {Roncin}},\ and\ \bibinfo {author}
  {\bibfnamefont {M.}~\bibnamefont {Arndt}},\ }\bibfield  {title} {\bibinfo
  {title} {Coherent diffraction of hydrogen through the 246 pm lattice of
  graphene},\ }\href {https://doi.org/10.1088/1367-2630/ab05ed} {\bibfield
  {journal} {\bibinfo  {journal} {New Journal of Physics}\ }\textbf {\bibinfo
  {volume} {21}},\ \bibinfo {pages} {033004} (\bibinfo {year}
  {2019})}\BibitemShut {NoStop}%
\bibitem [{\citenamefont {Kanitz}\ \emph {et~al.}(2025)\citenamefont {Kanitz},
  \citenamefont {Bühler}, \citenamefont {Zobač}, \citenamefont {Robinson},
  \citenamefont {Susi}, \citenamefont {Debiossac},\ and\ \citenamefont
  {Brand}}]{Kanitz2025}%
  \BibitemOpen
  \bibfield  {author} {\bibinfo {author} {\bibfnamefont {C.}~\bibnamefont
  {Kanitz}}, \bibinfo {author} {\bibfnamefont {J.}~\bibnamefont {Bühler}},
  \bibinfo {author} {\bibfnamefont {V.}~\bibnamefont {Zobač}}, \bibinfo
  {author} {\bibfnamefont {J.~J.}\ \bibnamefont {Robinson}}, \bibinfo {author}
  {\bibfnamefont {T.}~\bibnamefont {Susi}}, \bibinfo {author} {\bibfnamefont
  {M.}~\bibnamefont {Debiossac}},\ and\ \bibinfo {author} {\bibfnamefont
  {C.}~\bibnamefont {Brand}},\ }\bibfield  {title} {\bibinfo {title}
  {Diffraction of helium and hydrogen atoms through single-layer graphene},\
  }\href {https://doi.org/10.1126/science.adx5679} {\bibfield  {journal}
  {\bibinfo  {journal} {Science}\ }\textbf {\bibinfo {volume} {389}},\ \bibinfo
  {pages} {724} (\bibinfo {year} {2025})}\BibitemShut {NoStop}%
\bibitem [{\citenamefont {Labaigt}\ \emph {et~al.}(2014)\citenamefont
  {Labaigt}, \citenamefont {Dubois},\ and\ \citenamefont
  {Hansen}}]{Labaigt2014}%
  \BibitemOpen
  \bibfield  {author} {\bibinfo {author} {\bibfnamefont {G.}~\bibnamefont
  {Labaigt}}, \bibinfo {author} {\bibfnamefont {A.}~\bibnamefont {Dubois}},\
  and\ \bibinfo {author} {\bibfnamefont {J.~P.}\ \bibnamefont {Hansen}},\
  }\bibfield  {title} {\bibinfo {title} {Electron capture imaging of
  two-dimensional materials},\ }\href
  {https://doi.org/10.1103/PhysRevB.89.245438} {\bibfield  {journal} {\bibinfo
  {journal} {Phys. Rev. B}\ }\textbf {\bibinfo {volume} {89}},\ \bibinfo
  {pages} {245438} (\bibinfo {year} {2014})}\BibitemShut {NoStop}%
\bibitem [{sup(2025)}]{supplemental}%
  \BibitemOpen
  \href@noop {} {} (\bibinfo {year} {2025}),\ \bibinfo {note} {see Supplemental
  Material for further information on the experimental setup and method,
  graphene sample characterization, energy dependence of diffraction and energy
  loss, Debye-Waller factor, and \textsc{Quantum ESPRESSO}
  parameters}\BibitemShut {NoStop}%
\bibitem [{\citenamefont {Tian}\ \emph {et~al.}(2016)\citenamefont {Tian},
  \citenamefont {Yang}, \citenamefont {Liu}, \citenamefont {Wang},
  \citenamefont {Pan}, \citenamefont {Gu},\ and\ \citenamefont
  {Li}}]{Tian2016}%
  \BibitemOpen
  \bibfield  {author} {\bibinfo {author} {\bibfnamefont {S.}~\bibnamefont
  {Tian}}, \bibinfo {author} {\bibfnamefont {Y.}~\bibnamefont {Yang}}, \bibinfo
  {author} {\bibfnamefont {Z.}~\bibnamefont {Liu}}, \bibinfo {author}
  {\bibfnamefont {C.}~\bibnamefont {Wang}}, \bibinfo {author} {\bibfnamefont
  {R.}~\bibnamefont {Pan}}, \bibinfo {author} {\bibfnamefont {C.}~\bibnamefont
  {Gu}},\ and\ \bibinfo {author} {\bibfnamefont {J.}~\bibnamefont {Li}},\
  }\bibfield  {title} {\bibinfo {title} {Temperature-dependent {Raman}
  investigation on suspended graphene: Contribution from thermal expansion
  coefficient mismatch between graphene and substrate},\ }\href
  {https://doi.org/10.1016/j.carbon.2016.03.046} {\bibfield  {journal}
  {\bibinfo  {journal} {Carbon}\ }\textbf {\bibinfo {volume} {104}},\ \bibinfo
  {pages} {27–32} (\bibinfo {year} {2016})}\BibitemShut {NoStop}%
\bibitem [{\citenamefont {Meyer}\ \emph {et~al.}(2007)\citenamefont {Meyer},
  \citenamefont {Geim}, \citenamefont {Katsnelson}, \citenamefont {Novoselov},
  \citenamefont {Booth},\ and\ \citenamefont {Roth}}]{Meyer2007}%
  \BibitemOpen
  \bibfield  {author} {\bibinfo {author} {\bibfnamefont {J.~C.}\ \bibnamefont
  {Meyer}}, \bibinfo {author} {\bibfnamefont {A.~K.}\ \bibnamefont {Geim}},
  \bibinfo {author} {\bibfnamefont {M.~I.}\ \bibnamefont {Katsnelson}},
  \bibinfo {author} {\bibfnamefont {K.~S.}\ \bibnamefont {Novoselov}}, \bibinfo
  {author} {\bibfnamefont {T.~J.}\ \bibnamefont {Booth}},\ and\ \bibinfo
  {author} {\bibfnamefont {S.}~\bibnamefont {Roth}},\ }\bibfield  {title}
  {\bibinfo {title} {The structure of suspended graphene sheets},\ }\href
  {https://doi.org/10.1038/nature05545} {\bibfield  {journal} {\bibinfo
  {journal} {Nature}\ }\textbf {\bibinfo {volume} {446}},\ \bibinfo {pages}
  {60–63} (\bibinfo {year} {2007})}\BibitemShut {NoStop}%
\bibitem [{\citenamefont {Bao}\ \emph {et~al.}(2009)\citenamefont {Bao},
  \citenamefont {Miao}, \citenamefont {Chen}, \citenamefont {Zhang},
  \citenamefont {Jang}, \citenamefont {Dames},\ and\ \citenamefont
  {Lau}}]{Bao2009}%
  \BibitemOpen
  \bibfield  {author} {\bibinfo {author} {\bibfnamefont {W.}~\bibnamefont
  {Bao}}, \bibinfo {author} {\bibfnamefont {F.}~\bibnamefont {Miao}}, \bibinfo
  {author} {\bibfnamefont {Z.}~\bibnamefont {Chen}}, \bibinfo {author}
  {\bibfnamefont {H.}~\bibnamefont {Zhang}}, \bibinfo {author} {\bibfnamefont
  {W.}~\bibnamefont {Jang}}, \bibinfo {author} {\bibfnamefont {C.}~\bibnamefont
  {Dames}},\ and\ \bibinfo {author} {\bibfnamefont {C.~N.}\ \bibnamefont
  {Lau}},\ }\bibfield  {title} {\bibinfo {title} {Controlled ripple texturing
  of suspended graphene and ultrathin graphite membranes},\ }\href
  {https://doi.org/10.1038/nnano.2009.191} {\bibfield  {journal} {\bibinfo
  {journal} {Nature Nanotechnology}\ }\textbf {\bibinfo {volume} {4}},\
  \bibinfo {pages} {562–566} (\bibinfo {year} {2009})}\BibitemShut {NoStop}%
\bibitem [{\citenamefont {Niggas}\ \emph {et~al.}(2020)\citenamefont {Niggas},
  \citenamefont {Schwestka}, \citenamefont {Creutzburg}, \citenamefont {Gupta},
  \citenamefont {Eder}, \citenamefont {Bayer}, \citenamefont {Aumayr},\ and\
  \citenamefont {Wilhelm}}]{Niggas2020}%
  \BibitemOpen
  \bibfield  {author} {\bibinfo {author} {\bibfnamefont {A.}~\bibnamefont
  {Niggas}}, \bibinfo {author} {\bibfnamefont {J.}~\bibnamefont {Schwestka}},
  \bibinfo {author} {\bibfnamefont {S.}~\bibnamefont {Creutzburg}}, \bibinfo
  {author} {\bibfnamefont {T.}~\bibnamefont {Gupta}}, \bibinfo {author}
  {\bibfnamefont {D.}~\bibnamefont {Eder}}, \bibinfo {author} {\bibfnamefont
  {B.~C.}\ \bibnamefont {Bayer}}, \bibinfo {author} {\bibfnamefont
  {F.}~\bibnamefont {Aumayr}},\ and\ \bibinfo {author} {\bibfnamefont {R.~A.}\
  \bibnamefont {Wilhelm}},\ }\bibfield  {title} {\bibinfo {title} {The role of
  contaminations in ion beam spectroscopy with freestanding {2D} materials: A
  study on thermal treatment},\ }\href {https://doi.org/10.1063/5.0011255}
  {\bibfield  {journal} {\bibinfo  {journal} {The Journal of Chemical Physics}\
  }\textbf {\bibinfo {volume} {153}},\ \bibinfo {pages} {014702} (\bibinfo
  {year} {2020})}\BibitemShut {NoStop}%
\bibitem [{\citenamefont {Ferrah}\ \emph {et~al.}(2016)\citenamefont {Ferrah},
  \citenamefont {Renault}, \citenamefont {Petit-Etienne}, \citenamefont
  {Okuno}, \citenamefont {Berne}, \citenamefont {Bouchiat},\ and\ \citenamefont
  {Cunge}}]{Ferrah2016}%
  \BibitemOpen
  \bibfield  {author} {\bibinfo {author} {\bibfnamefont {D.}~\bibnamefont
  {Ferrah}}, \bibinfo {author} {\bibfnamefont {O.}~\bibnamefont {Renault}},
  \bibinfo {author} {\bibfnamefont {C.}~\bibnamefont {Petit-Etienne}}, \bibinfo
  {author} {\bibfnamefont {H.}~\bibnamefont {Okuno}}, \bibinfo {author}
  {\bibfnamefont {C.}~\bibnamefont {Berne}}, \bibinfo {author} {\bibfnamefont
  {V.}~\bibnamefont {Bouchiat}},\ and\ \bibinfo {author} {\bibfnamefont
  {G.}~\bibnamefont {Cunge}},\ }\bibfield  {title} {\bibinfo {title} {{XPS}
  investigations of graphene surface cleaning using {H$_2$-} and {Cl$_2$-}based
  inductively coupled plasma},\ }\href
  {https://doi.org/https://doi.org/10.1002/sia.6010} {\bibfield  {journal}
  {\bibinfo  {journal} {Surface and Interface Analysis}\ }\textbf {\bibinfo
  {volume} {48}},\ \bibinfo {pages} {451} (\bibinfo {year} {2016})}\BibitemShut
  {NoStop}%
\bibitem [{\citenamefont {Whitener}\ \emph {et~al.}(2014)\citenamefont
  {Whitener}, \citenamefont {Lee}, \citenamefont {Campbell}, \citenamefont
  {Robinson},\ and\ \citenamefont {Sheehan}}]{Whitener2014}%
  \BibitemOpen
  \bibfield  {author} {\bibinfo {author} {\bibfnamefont {K.~E.}\ \bibnamefont
  {Whitener}}, \bibinfo {author} {\bibfnamefont {W.~K.}\ \bibnamefont {Lee}},
  \bibinfo {author} {\bibfnamefont {P.~M.}\ \bibnamefont {Campbell}}, \bibinfo
  {author} {\bibfnamefont {J.~T.}\ \bibnamefont {Robinson}},\ and\ \bibinfo
  {author} {\bibfnamefont {P.~E.}\ \bibnamefont {Sheehan}},\ }\bibfield
  {title} {\bibinfo {title} {Chemical hydrogenation of single-layer graphene
  enables completely reversible removal of electrical conductivity},\ }\href
  {https://doi.org/https://doi.org/10.1016/j.carbon.2014.02.022} {\bibfield
  {journal} {\bibinfo  {journal} {Carbon}\ }\textbf {\bibinfo {volume} {72}},\
  \bibinfo {pages} {348} (\bibinfo {year} {2014})}\BibitemShut {NoStop}%
\bibitem [{\citenamefont {Zhao}\ \emph {et~al.}(2017)\citenamefont {Zhao},
  \citenamefont {Xia}, \citenamefont {Lin}, \citenamefont {Xiao}, \citenamefont
  {Liu}, \citenamefont {Lin}, \citenamefont {Peng}, \citenamefont {Zhu},
  \citenamefont {Yu}, \citenamefont {Lei}, \citenamefont {Wang}, \citenamefont
  {Zhang}, \citenamefont {Xu}, \citenamefont {Zhao}, \citenamefont {Peng},
  \citenamefont {Li}, \citenamefont {Duan}, \citenamefont {Liu}, \citenamefont
  {Fan},\ and\ \citenamefont {Jiang}}]{Zhao2017}%
  \BibitemOpen
  \bibfield  {author} {\bibinfo {author} {\bibfnamefont {W.}~\bibnamefont
  {Zhao}}, \bibinfo {author} {\bibfnamefont {B.}~\bibnamefont {Xia}}, \bibinfo
  {author} {\bibfnamefont {L.}~\bibnamefont {Lin}}, \bibinfo {author}
  {\bibfnamefont {X.}~\bibnamefont {Xiao}}, \bibinfo {author} {\bibfnamefont
  {P.}~\bibnamefont {Liu}}, \bibinfo {author} {\bibfnamefont {X.}~\bibnamefont
  {Lin}}, \bibinfo {author} {\bibfnamefont {H.}~\bibnamefont {Peng}}, \bibinfo
  {author} {\bibfnamefont {Y.}~\bibnamefont {Zhu}}, \bibinfo {author}
  {\bibfnamefont {R.}~\bibnamefont {Yu}}, \bibinfo {author} {\bibfnamefont
  {P.}~\bibnamefont {Lei}}, \bibinfo {author} {\bibfnamefont {J.}~\bibnamefont
  {Wang}}, \bibinfo {author} {\bibfnamefont {L.}~\bibnamefont {Zhang}},
  \bibinfo {author} {\bibfnamefont {Y.}~\bibnamefont {Xu}}, \bibinfo {author}
  {\bibfnamefont {M.}~\bibnamefont {Zhao}}, \bibinfo {author} {\bibfnamefont
  {L.}~\bibnamefont {Peng}}, \bibinfo {author} {\bibfnamefont {Q.}~\bibnamefont
  {Li}}, \bibinfo {author} {\bibfnamefont {W.}~\bibnamefont {Duan}}, \bibinfo
  {author} {\bibfnamefont {Z.}~\bibnamefont {Liu}}, \bibinfo {author}
  {\bibfnamefont {S.}~\bibnamefont {Fan}},\ and\ \bibinfo {author}
  {\bibfnamefont {K.}~\bibnamefont {Jiang}},\ }\bibfield  {title} {\bibinfo
  {title} {Low-energy transmission electron diffraction and imaging of
  large-area graphene},\ }\href {https://doi.org/10.1126/sciadv.1603231}
  {\bibfield  {journal} {\bibinfo  {journal} {Science Advances}\ }\textbf
  {\bibinfo {volume} {3}},\ \bibinfo {pages} {e1603231} (\bibinfo {year}
  {2017})}\BibitemShut {NoStop}%
\bibitem [{\citenamefont {Neubeck}\ \emph {et~al.}(2010)\citenamefont
  {Neubeck}, \citenamefont {You}, \citenamefont {Ni}, \citenamefont {Blake},
  \citenamefont {Shen}, \citenamefont {Geim},\ and\ \citenamefont
  {Novoselov}}]{Neubeck2010}%
  \BibitemOpen
  \bibfield  {author} {\bibinfo {author} {\bibfnamefont {S.}~\bibnamefont
  {Neubeck}}, \bibinfo {author} {\bibfnamefont {Y.~M.}\ \bibnamefont {You}},
  \bibinfo {author} {\bibfnamefont {Z.~H.}\ \bibnamefont {Ni}}, \bibinfo
  {author} {\bibfnamefont {P.}~\bibnamefont {Blake}}, \bibinfo {author}
  {\bibfnamefont {Z.~X.}\ \bibnamefont {Shen}}, \bibinfo {author}
  {\bibfnamefont {A.~K.}\ \bibnamefont {Geim}},\ and\ \bibinfo {author}
  {\bibfnamefont {K.~S.}\ \bibnamefont {Novoselov}},\ }\bibfield  {title}
  {\bibinfo {title} {Direct determination of the crystallographic orientation
  of graphene edges by atomic resolution imaging},\ }\href
  {https://doi.org/10.1063/1.3467468} {\bibfield  {journal} {\bibinfo
  {journal} {Applied Physics Letters}\ }\textbf {\bibinfo {volume} {97}},\
  \bibinfo {pages} {053110} (\bibinfo {year} {2010})}\BibitemShut {NoStop}%
\bibitem [{\citenamefont {Caplins}\ \emph {et~al.}(2019)\citenamefont
  {Caplins}, \citenamefont {Holm},\ and\ \citenamefont {Keller}}]{Caplins2019}%
  \BibitemOpen
  \bibfield  {author} {\bibinfo {author} {\bibfnamefont {B.~W.}\ \bibnamefont
  {Caplins}}, \bibinfo {author} {\bibfnamefont {J.~D.}\ \bibnamefont {Holm}},\
  and\ \bibinfo {author} {\bibfnamefont {R.~R.}\ \bibnamefont {Keller}},\
  }\bibfield  {title} {\bibinfo {title} {Orientation mapping of graphene in a
  scanning electron microscope},\ }\href
  {https://doi.org/https://doi.org/10.1016/j.carbon.2019.04.042} {\bibfield
  {journal} {\bibinfo  {journal} {Carbon}\ }\textbf {\bibinfo {volume} {149}},\
  \bibinfo {pages} {400} (\bibinfo {year} {2019})}\BibitemShut {NoStop}%
\bibitem [{\citenamefont {Huang}\ \emph {et~al.}(2011)\citenamefont {Huang},
  \citenamefont {Ruiz-Vargas}, \citenamefont {van~der Zande}, \citenamefont
  {Whitney}, \citenamefont {Levendorf}, \citenamefont {Kevek}, \citenamefont
  {Garg}, \citenamefont {Alden}, \citenamefont {Hustedt}, \citenamefont {Zhu},
  \citenamefont {Park}, \citenamefont {McEuen},\ and\ \citenamefont
  {Muller}}]{Huang2011}%
  \BibitemOpen
  \bibfield  {author} {\bibinfo {author} {\bibfnamefont {P.~Y.}\ \bibnamefont
  {Huang}}, \bibinfo {author} {\bibfnamefont {C.~S.}\ \bibnamefont
  {Ruiz-Vargas}}, \bibinfo {author} {\bibfnamefont {A.~M.}\ \bibnamefont
  {van~der Zande}}, \bibinfo {author} {\bibfnamefont {W.~S.}\ \bibnamefont
  {Whitney}}, \bibinfo {author} {\bibfnamefont {M.~P.}\ \bibnamefont
  {Levendorf}}, \bibinfo {author} {\bibfnamefont {J.~W.}\ \bibnamefont
  {Kevek}}, \bibinfo {author} {\bibfnamefont {S.}~\bibnamefont {Garg}},
  \bibinfo {author} {\bibfnamefont {J.~S.}\ \bibnamefont {Alden}}, \bibinfo
  {author} {\bibfnamefont {C.~J.}\ \bibnamefont {Hustedt}}, \bibinfo {author}
  {\bibfnamefont {Y.}~\bibnamefont {Zhu}}, \bibinfo {author} {\bibfnamefont
  {J.}~\bibnamefont {Park}}, \bibinfo {author} {\bibfnamefont {P.~L.}\
  \bibnamefont {McEuen}},\ and\ \bibinfo {author} {\bibfnamefont {D.~A.}\
  \bibnamefont {Muller}},\ }\bibfield  {title} {\bibinfo {title} {Grains and
  grain boundaries in single-layer graphene atomic patchwork quilts},\ }\href
  {https://doi.org/10.1038/nature09718} {\bibfield  {journal} {\bibinfo
  {journal} {Nature}\ }\textbf {\bibinfo {volume} {469}},\ \bibinfo {pages}
  {389} (\bibinfo {year} {2011})}\BibitemShut {NoStop}%
\bibitem [{\citenamefont {Shevitski}\ \emph {et~al.}(2013)\citenamefont
  {Shevitski}, \citenamefont {Mecklenburg}, \citenamefont {Hubbard},
  \citenamefont {White}, \citenamefont {Dawson}, \citenamefont {Lodge},
  \citenamefont {Ishigami},\ and\ \citenamefont {Regan}}]{Shevitski2013}%
  \BibitemOpen
  \bibfield  {author} {\bibinfo {author} {\bibfnamefont {B.}~\bibnamefont
  {Shevitski}}, \bibinfo {author} {\bibfnamefont {M.}~\bibnamefont
  {Mecklenburg}}, \bibinfo {author} {\bibfnamefont {W.~A.}\ \bibnamefont
  {Hubbard}}, \bibinfo {author} {\bibfnamefont {E.~R.}\ \bibnamefont {White}},
  \bibinfo {author} {\bibfnamefont {B.}~\bibnamefont {Dawson}}, \bibinfo
  {author} {\bibfnamefont {M.~S.}\ \bibnamefont {Lodge}}, \bibinfo {author}
  {\bibfnamefont {M.}~\bibnamefont {Ishigami}},\ and\ \bibinfo {author}
  {\bibfnamefont {B.~C.}\ \bibnamefont {Regan}},\ }\bibfield  {title} {\bibinfo
  {title} {Dark-field transmission electron microscopy and the {Debye}-{Waller}
  factor of graphene},\ }\href {https://doi.org/10.1103/PhysRevB.87.045417}
  {\bibfield  {journal} {\bibinfo  {journal} {Phys. Rev. B}\ }\textbf {\bibinfo
  {volume} {87}},\ \bibinfo {pages} {045417} (\bibinfo {year}
  {2013})}\BibitemShut {NoStop}%
\bibitem [{\citenamefont {Landau}\ and\ \citenamefont
  {Lifshitz}(1981)}]{Landau1981}%
  \BibitemOpen
  \bibfield  {author} {\bibinfo {author} {\bibfnamefont {L.}~\bibnamefont
  {Landau}}\ and\ \bibinfo {author} {\bibfnamefont {E.}~\bibnamefont
  {Lifshitz}},\ }\href@noop {} {\emph {\bibinfo {title} {{Quantum Mechanics:
  Non-Relativistic Theory}}}},\ Vol.~\bibinfo {volume} {3}\ (\bibinfo
  {publisher} {Elsevier},\ \bibinfo {year} {1981})\ p.\ \bibinfo {pages}
  {160}\BibitemShut {NoStop}%
\bibitem [{\citenamefont {Ehemann}\ \emph {et~al.}(2012)\citenamefont
  {Ehemann}, \citenamefont {Krsti{\'{c}}}, \citenamefont {Dadras},
  \citenamefont {Kent},\ and\ \citenamefont {Jakowski}}]{Ehemann2012}%
  \BibitemOpen
  \bibfield  {author} {\bibinfo {author} {\bibfnamefont {R.~C.}\ \bibnamefont
  {Ehemann}}, \bibinfo {author} {\bibfnamefont {P.~S.}\ \bibnamefont
  {Krsti{\'{c}}}}, \bibinfo {author} {\bibfnamefont {J.}~\bibnamefont
  {Dadras}}, \bibinfo {author} {\bibfnamefont {P.~R.}\ \bibnamefont {Kent}},\
  and\ \bibinfo {author} {\bibfnamefont {J.}~\bibnamefont {Jakowski}},\
  }\bibfield  {title} {\bibinfo {title} {Detection of hydrogen using
  graphene},\ }\href {https://doi.org/10.1186/1556-276X-7-198} {\bibfield
  {journal} {\bibinfo  {journal} {Nanoscale Research Letters}\ }\textbf
  {\bibinfo {volume} {7}},\ \bibinfo {pages} {198} (\bibinfo {year}
  {2012})}\BibitemShut {NoStop}%
\bibitem [{\citenamefont {Newton}(2013)}]{Newton1982}%
  \BibitemOpen
  \bibfield  {author} {\bibinfo {author} {\bibfnamefont {R.~G.}\ \bibnamefont
  {Newton}},\ }\href@noop {} {\emph {\bibinfo {title} {Scattering theory of
  waves and particles}}}\ (\bibinfo  {publisher} {Springer Science \& Business
  Media},\ \bibinfo {year} {2013})\BibitemShut {NoStop}%
\bibitem [{\citenamefont {Giannozzi}\ \emph {et~al.}(2009)\citenamefont
  {Giannozzi}, \citenamefont {Baroni}, \citenamefont {Bonini}, \citenamefont
  {Calandra}, \citenamefont {Car}, \citenamefont {Cavazzoni}, \citenamefont
  {Ceresoli}, \citenamefont {Chiarotti}, \citenamefont {Cococcioni},
  \citenamefont {Dabo}, \citenamefont {Dal~Corso}, \citenamefont
  {de~Gironcoli}, \citenamefont {Fabris}, \citenamefont {Fratesi},
  \citenamefont {Gebauer}, \citenamefont {Gerstmann}, \citenamefont
  {Gougoussis}, \citenamefont {Kokalj}, \citenamefont {Lazzeri}, \citenamefont
  {Martin-Samos}, \citenamefont {Marzari}, \citenamefont {Mauri}, \citenamefont
  {Mazzarello}, \citenamefont {Paolini}, \citenamefont {Pasquarello},
  \citenamefont {Paulatto}, \citenamefont {Sbraccia}, \citenamefont {Scandolo},
  \citenamefont {Sclauzero}, \citenamefont {Seitsonen}, \citenamefont
  {Smogunov}, \citenamefont {Umari},\ and\ \citenamefont
  {Wentzcovitch}}]{QE-2009}%
  \BibitemOpen
  \bibfield  {author} {\bibinfo {author} {\bibfnamefont {P.}~\bibnamefont
  {Giannozzi}}, \bibinfo {author} {\bibfnamefont {S.}~\bibnamefont {Baroni}},
  \bibinfo {author} {\bibfnamefont {N.}~\bibnamefont {Bonini}}, \bibinfo
  {author} {\bibfnamefont {M.}~\bibnamefont {Calandra}}, \bibinfo {author}
  {\bibfnamefont {R.}~\bibnamefont {Car}}, \bibinfo {author} {\bibfnamefont
  {C.}~\bibnamefont {Cavazzoni}}, \bibinfo {author} {\bibfnamefont
  {D.}~\bibnamefont {Ceresoli}}, \bibinfo {author} {\bibfnamefont {G.~L.}\
  \bibnamefont {Chiarotti}}, \bibinfo {author} {\bibfnamefont {M.}~\bibnamefont
  {Cococcioni}}, \bibinfo {author} {\bibfnamefont {I.}~\bibnamefont {Dabo}},
  \bibinfo {author} {\bibfnamefont {A.}~\bibnamefont {Dal~Corso}}, \bibinfo
  {author} {\bibfnamefont {S.}~\bibnamefont {de~Gironcoli}}, \bibinfo {author}
  {\bibfnamefont {S.}~\bibnamefont {Fabris}}, \bibinfo {author} {\bibfnamefont
  {G.}~\bibnamefont {Fratesi}}, \bibinfo {author} {\bibfnamefont
  {R.}~\bibnamefont {Gebauer}}, \bibinfo {author} {\bibfnamefont
  {U.}~\bibnamefont {Gerstmann}}, \bibinfo {author} {\bibfnamefont
  {C.}~\bibnamefont {Gougoussis}}, \bibinfo {author} {\bibfnamefont
  {A.}~\bibnamefont {Kokalj}}, \bibinfo {author} {\bibfnamefont
  {M.}~\bibnamefont {Lazzeri}}, \bibinfo {author} {\bibfnamefont
  {L.}~\bibnamefont {Martin-Samos}}, \bibinfo {author} {\bibfnamefont
  {N.}~\bibnamefont {Marzari}}, \bibinfo {author} {\bibfnamefont
  {F.}~\bibnamefont {Mauri}}, \bibinfo {author} {\bibfnamefont
  {R.}~\bibnamefont {Mazzarello}}, \bibinfo {author} {\bibfnamefont
  {S.}~\bibnamefont {Paolini}}, \bibinfo {author} {\bibfnamefont
  {A.}~\bibnamefont {Pasquarello}}, \bibinfo {author} {\bibfnamefont
  {L.}~\bibnamefont {Paulatto}}, \bibinfo {author} {\bibfnamefont
  {C.}~\bibnamefont {Sbraccia}}, \bibinfo {author} {\bibfnamefont
  {S.}~\bibnamefont {Scandolo}}, \bibinfo {author} {\bibfnamefont
  {G.}~\bibnamefont {Sclauzero}}, \bibinfo {author} {\bibfnamefont {A.~P.}\
  \bibnamefont {Seitsonen}}, \bibinfo {author} {\bibfnamefont {A.}~\bibnamefont
  {Smogunov}}, \bibinfo {author} {\bibfnamefont {P.}~\bibnamefont {Umari}},\
  and\ \bibinfo {author} {\bibfnamefont {R.~M.}\ \bibnamefont {Wentzcovitch}},\
  }\bibfield  {title} {\bibinfo {title} {{QUANTUM ESPRESSO}: a modular and
  open-source software project for quantum simulations of materials},\ }\href
  {https://doi.org/10.1088/0953-8984/21/39/395502} {\bibfield  {journal}
  {\bibinfo  {journal} {Journal of Physics: Condensed Matter}\ }\textbf
  {\bibinfo {volume} {21}},\ \bibinfo {pages} {395502} (\bibinfo {year}
  {2009})}\BibitemShut {NoStop}%
\bibitem [{\citenamefont {Giannozzi}\ \emph {et~al.}(2017)\citenamefont
  {Giannozzi}, \citenamefont {Andreussi}, \citenamefont {Brumme}, \citenamefont
  {Bunau}, \citenamefont {Buongiorno~Nardelli}, \citenamefont {Calandra},
  \citenamefont {Car}, \citenamefont {Cavazzoni}, \citenamefont {Ceresoli},
  \citenamefont {Cococcioni}, \citenamefont {Colonna}, \citenamefont
  {Carnimeo}, \citenamefont {Dal~Corso}, \citenamefont {de~Gironcoli},
  \citenamefont {Delugas}, \citenamefont {DiStasio}, \citenamefont {Ferretti},
  \citenamefont {Floris}, \citenamefont {Fratesi}, \citenamefont {Fugallo},
  \citenamefont {Gebauer}, \citenamefont {Gerstmann}, \citenamefont {Giustino},
  \citenamefont {Gorni}, \citenamefont {Jia}, \citenamefont {Kawamura},
  \citenamefont {Ko}, \citenamefont {Kokalj}, \citenamefont {Küçükbenli},
  \citenamefont {Lazzeri}, \citenamefont {Marsili}, \citenamefont {Marzari},
  \citenamefont {Mauri}, \citenamefont {Nguyen}, \citenamefont {Nguyen},
  \citenamefont {Otero-de-la Roza}, \citenamefont {Paulatto}, \citenamefont
  {Poncé}, \citenamefont {Rocca}, \citenamefont {Sabatini}, \citenamefont
  {Santra}, \citenamefont {Schlipf}, \citenamefont {Seitsonen}, \citenamefont
  {Smogunov}, \citenamefont {Timrov}, \citenamefont {Thonhauser}, \citenamefont
  {Umari}, \citenamefont {Vast}, \citenamefont {Wu},\ and\ \citenamefont
  {Baroni}}]{QE-2017}%
  \BibitemOpen
  \bibfield  {author} {\bibinfo {author} {\bibfnamefont {P.}~\bibnamefont
  {Giannozzi}}, \bibinfo {author} {\bibfnamefont {O.}~\bibnamefont
  {Andreussi}}, \bibinfo {author} {\bibfnamefont {T.}~\bibnamefont {Brumme}},
  \bibinfo {author} {\bibfnamefont {O.}~\bibnamefont {Bunau}}, \bibinfo
  {author} {\bibfnamefont {M.}~\bibnamefont {Buongiorno~Nardelli}}, \bibinfo
  {author} {\bibfnamefont {M.}~\bibnamefont {Calandra}}, \bibinfo {author}
  {\bibfnamefont {R.}~\bibnamefont {Car}}, \bibinfo {author} {\bibfnamefont
  {C.}~\bibnamefont {Cavazzoni}}, \bibinfo {author} {\bibfnamefont
  {D.}~\bibnamefont {Ceresoli}}, \bibinfo {author} {\bibfnamefont
  {M.}~\bibnamefont {Cococcioni}}, \bibinfo {author} {\bibfnamefont
  {N.}~\bibnamefont {Colonna}}, \bibinfo {author} {\bibfnamefont
  {I.}~\bibnamefont {Carnimeo}}, \bibinfo {author} {\bibfnamefont
  {A.}~\bibnamefont {Dal~Corso}}, \bibinfo {author} {\bibfnamefont
  {S.}~\bibnamefont {de~Gironcoli}}, \bibinfo {author} {\bibfnamefont
  {P.}~\bibnamefont {Delugas}}, \bibinfo {author} {\bibfnamefont {R.~A.}\
  \bibnamefont {DiStasio}}, \bibinfo {author} {\bibfnamefont {A.}~\bibnamefont
  {Ferretti}}, \bibinfo {author} {\bibfnamefont {A.}~\bibnamefont {Floris}},
  \bibinfo {author} {\bibfnamefont {G.}~\bibnamefont {Fratesi}}, \bibinfo
  {author} {\bibfnamefont {G.}~\bibnamefont {Fugallo}}, \bibinfo {author}
  {\bibfnamefont {R.}~\bibnamefont {Gebauer}}, \bibinfo {author} {\bibfnamefont
  {U.}~\bibnamefont {Gerstmann}}, \bibinfo {author} {\bibfnamefont
  {F.}~\bibnamefont {Giustino}}, \bibinfo {author} {\bibfnamefont
  {T.}~\bibnamefont {Gorni}}, \bibinfo {author} {\bibfnamefont
  {J.}~\bibnamefont {Jia}}, \bibinfo {author} {\bibfnamefont {M.}~\bibnamefont
  {Kawamura}}, \bibinfo {author} {\bibfnamefont {H.-Y.}\ \bibnamefont {Ko}},
  \bibinfo {author} {\bibfnamefont {A.}~\bibnamefont {Kokalj}}, \bibinfo
  {author} {\bibfnamefont {E.}~\bibnamefont {Küçükbenli}}, \bibinfo {author}
  {\bibfnamefont {M.}~\bibnamefont {Lazzeri}}, \bibinfo {author} {\bibfnamefont
  {M.}~\bibnamefont {Marsili}}, \bibinfo {author} {\bibfnamefont
  {N.}~\bibnamefont {Marzari}}, \bibinfo {author} {\bibfnamefont
  {F.}~\bibnamefont {Mauri}}, \bibinfo {author} {\bibfnamefont {N.~L.}\
  \bibnamefont {Nguyen}}, \bibinfo {author} {\bibfnamefont {H.-V.}\
  \bibnamefont {Nguyen}}, \bibinfo {author} {\bibfnamefont {A.}~\bibnamefont
  {Otero-de-la Roza}}, \bibinfo {author} {\bibfnamefont {L.}~\bibnamefont
  {Paulatto}}, \bibinfo {author} {\bibfnamefont {S.}~\bibnamefont {Poncé}},
  \bibinfo {author} {\bibfnamefont {D.}~\bibnamefont {Rocca}}, \bibinfo
  {author} {\bibfnamefont {R.}~\bibnamefont {Sabatini}}, \bibinfo {author}
  {\bibfnamefont {B.}~\bibnamefont {Santra}}, \bibinfo {author} {\bibfnamefont
  {M.}~\bibnamefont {Schlipf}}, \bibinfo {author} {\bibfnamefont {A.~P.}\
  \bibnamefont {Seitsonen}}, \bibinfo {author} {\bibfnamefont {A.}~\bibnamefont
  {Smogunov}}, \bibinfo {author} {\bibfnamefont {I.}~\bibnamefont {Timrov}},
  \bibinfo {author} {\bibfnamefont {T.}~\bibnamefont {Thonhauser}}, \bibinfo
  {author} {\bibfnamefont {P.}~\bibnamefont {Umari}}, \bibinfo {author}
  {\bibfnamefont {N.}~\bibnamefont {Vast}}, \bibinfo {author} {\bibfnamefont
  {X.}~\bibnamefont {Wu}},\ and\ \bibinfo {author} {\bibfnamefont
  {S.}~\bibnamefont {Baroni}},\ }\bibfield  {title} {\bibinfo {title} {Advanced
  capabilities for materials modelling with {Quantum ESPRESSO}},\ }\href
  {https://doi.org/10.1088/1361-648X/aa8f79} {\bibfield  {journal} {\bibinfo
  {journal} {Journal of Physics: Condensed Matter}\ }\textbf {\bibinfo {volume}
  {29}},\ \bibinfo {pages} {465901} (\bibinfo {year} {2017})}\BibitemShut
  {NoStop}%
\bibitem [{\citenamefont {Giannozzi}\ \emph {et~al.}(2020)\citenamefont
  {Giannozzi}, \citenamefont {Baseggio}, \citenamefont {Bonfà}, \citenamefont
  {Brunato}, \citenamefont {Car}, \citenamefont {Carnimeo}, \citenamefont
  {Cavazzoni}, \citenamefont {de~Gironcoli}, \citenamefont {Delugas},
  \citenamefont {Ferrari~Ruffino}, \citenamefont {Ferretti}, \citenamefont
  {Marzari}, \citenamefont {Timrov}, \citenamefont {Urru},\ and\ \citenamefont
  {Baroni}}]{QE-2020}%
  \BibitemOpen
  \bibfield  {author} {\bibinfo {author} {\bibfnamefont {P.}~\bibnamefont
  {Giannozzi}}, \bibinfo {author} {\bibfnamefont {O.}~\bibnamefont {Baseggio}},
  \bibinfo {author} {\bibfnamefont {P.}~\bibnamefont {Bonfà}}, \bibinfo
  {author} {\bibfnamefont {D.}~\bibnamefont {Brunato}}, \bibinfo {author}
  {\bibfnamefont {R.}~\bibnamefont {Car}}, \bibinfo {author} {\bibfnamefont
  {I.}~\bibnamefont {Carnimeo}}, \bibinfo {author} {\bibfnamefont
  {C.}~\bibnamefont {Cavazzoni}}, \bibinfo {author} {\bibfnamefont
  {S.}~\bibnamefont {de~Gironcoli}}, \bibinfo {author} {\bibfnamefont
  {P.}~\bibnamefont {Delugas}}, \bibinfo {author} {\bibfnamefont
  {F.}~\bibnamefont {Ferrari~Ruffino}}, \bibinfo {author} {\bibfnamefont
  {A.}~\bibnamefont {Ferretti}}, \bibinfo {author} {\bibfnamefont
  {N.}~\bibnamefont {Marzari}}, \bibinfo {author} {\bibfnamefont
  {I.}~\bibnamefont {Timrov}}, \bibinfo {author} {\bibfnamefont
  {A.}~\bibnamefont {Urru}},\ and\ \bibinfo {author} {\bibfnamefont
  {S.}~\bibnamefont {Baroni}},\ }\bibfield  {title} {\bibinfo {title} {{Quantum
  ESPRESSO} toward the exascale},\ }\href {https://doi.org/10.1063/5.0005082}
  {\bibfield  {journal} {\bibinfo  {journal} {The Journal of Chemical Physics}\
  }\textbf {\bibinfo {volume} {152}},\ \bibinfo {pages} {154105} (\bibinfo
  {year} {2020})}\BibitemShut {NoStop}%
\bibitem [{\citenamefont {Vermeeren}\ and\ \citenamefont
  {Bickelhaupt}(2023)}]{Vermeeren2022}%
  \BibitemOpen
  \bibfield  {author} {\bibinfo {author} {\bibfnamefont {P.}~\bibnamefont
  {Vermeeren}}\ and\ \bibinfo {author} {\bibfnamefont {F.~M.}\ \bibnamefont
  {Bickelhaupt}},\ }\bibfield  {title} {\bibinfo {title} {The abnormally long
  and weak methylidyne {C–H} bond},\ }\href
  {https://doi.org/https://doi.org/10.1002/ntls.20220039} {\bibfield  {journal}
  {\bibinfo  {journal} {Natural Sciences}\ }\textbf {\bibinfo {volume} {3}},\
  \bibinfo {pages} {e20220039} (\bibinfo {year} {2023})}\BibitemShut {NoStop}%
\bibitem [{\citenamefont {Elstner}\ \emph {et~al.}(1998)\citenamefont
  {Elstner}, \citenamefont {Porezag}, \citenamefont {Jungnickel}, \citenamefont
  {Elsner}, \citenamefont {Haugk}, \citenamefont {Frauenheim}, \citenamefont
  {Suhai},\ and\ \citenamefont {Seifert}}]{Elstner1998}%
  \BibitemOpen
  \bibfield  {author} {\bibinfo {author} {\bibfnamefont {M.}~\bibnamefont
  {Elstner}}, \bibinfo {author} {\bibfnamefont {D.}~\bibnamefont {Porezag}},
  \bibinfo {author} {\bibfnamefont {G.}~\bibnamefont {Jungnickel}}, \bibinfo
  {author} {\bibfnamefont {J.}~\bibnamefont {Elsner}}, \bibinfo {author}
  {\bibfnamefont {M.}~\bibnamefont {Haugk}}, \bibinfo {author} {\bibfnamefont
  {T.}~\bibnamefont {Frauenheim}}, \bibinfo {author} {\bibfnamefont
  {S.}~\bibnamefont {Suhai}},\ and\ \bibinfo {author} {\bibfnamefont
  {G.}~\bibnamefont {Seifert}},\ }\bibfield  {title} {\bibinfo {title}
  {Self-consistent-charge density-functional tight-binding method for
  simulations of complex materials properties},\ }\href
  {https://doi.org/10.1103/PhysRevB.58.7260} {\bibfield  {journal} {\bibinfo
  {journal} {Phys. Rev. B}\ }\textbf {\bibinfo {volume} {58}},\ \bibinfo
  {pages} {7260} (\bibinfo {year} {1998})}\BibitemShut {NoStop}%
\bibitem [{\citenamefont {Ivanovskaya}\ \emph {et~al.}(2010)\citenamefont
  {Ivanovskaya}, \citenamefont {Zobelli}, \citenamefont {Teillet-Billy},
  \citenamefont {Rougeau}, \citenamefont {Sidis},\ and\ \citenamefont
  {Briddon}}]{Ivanovskaya2010}%
  \BibitemOpen
  \bibfield  {author} {\bibinfo {author} {\bibfnamefont {V.~V.}\ \bibnamefont
  {Ivanovskaya}}, \bibinfo {author} {\bibfnamefont {A.}~\bibnamefont
  {Zobelli}}, \bibinfo {author} {\bibfnamefont {D.}~\bibnamefont
  {Teillet-Billy}}, \bibinfo {author} {\bibfnamefont {N.}~\bibnamefont
  {Rougeau}}, \bibinfo {author} {\bibfnamefont {V.}~\bibnamefont {Sidis}},\
  and\ \bibinfo {author} {\bibfnamefont {P.~R.}\ \bibnamefont {Briddon}},\
  }\bibfield  {title} {\bibinfo {title} {Hydrogen adsorption on graphene: a
  first principles study},\ }\href {https://doi.org/10.1140/epjb/e2010-00238-7}
  {\bibfield  {journal} {\bibinfo  {journal} {The European Physical Journal B}\
  }\textbf {\bibinfo {volume} {76}},\ \bibinfo {pages} {481} (\bibinfo {year}
  {2010})}\BibitemShut {NoStop}%
\end{thebibliography}%

\end{document}


\title{-- Supplemental Material -- \\ Fast hydrogen atom diffraction through monocrystalline graphene}

\author{Pierre Guichard}
\affiliation{Université de Strasbourg, CNRS, Institut de Physique et Chimie des Matériaux de Strasbourg, UMR 7504, 67000 Strasbourg, France}

\author{Arnaud Dochain}
\affiliation{Institute of Condensed Matter and Nanosciences, Universit\'e
	Catholique de Louvain, B-1348 Louvain-la-Neuve, Belgium}

\author{Rapha\"el Marion}
\affiliation{Institute of Condensed Matter and Nanosciences, Universit\'e
	Catholique de Louvain, B-1348 Louvain-la-Neuve, Belgium}
\affiliation{Royal Observatory of Belgium (ROB-ORB), B-1180 Brussels, Belgium}

\author{Pauline de Crombrugghe de Picquendaele}
\affiliation{Institute of Condensed Matter and Nanosciences, Universit\'e
	Catholique de Louvain, B-1348 Louvain-la-Neuve, Belgium}	

\author{Nicolas Lejeune}
\affiliation{Institute of Condensed Matter and Nanosciences, Universit\'e
	Catholique de Louvain, B-1348 Louvain-la-Neuve, Belgium}

\author{Beno\^it Hackens}
\affiliation{Institute of Condensed Matter and Nanosciences, Universit\'e
	Catholique de Louvain, B-1348 Louvain-la-Neuve, Belgium}	
\author{Paul-Antoine Hervieux}\email[]{paul-antoine.hervieux@ipcms.unistra.fr}
\affiliation{Université de Strasbourg, CNRS, Institut de Physique et Chimie des Matériaux de Strasbourg, UMR 7504, 67000 Strasbourg, France}

\author{Xavier Urbain}\email[]{xavier.urbain@uclouvain.be}
\affiliation{Institute of Condensed Matter and Nanosciences, Universit\'e
	Catholique de Louvain, B-1348 Louvain-la-Neuve, Belgium}

	\date{\today}
	
	\maketitle
	
\section{Experimental setup}

\begin{figure}[!h]
	\centering
	\includegraphics[width=0.7\columnwidth]{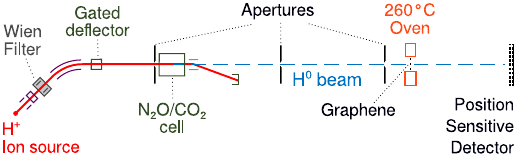}
	\caption{Schematics of the experimental setup.}
	\label{setup}
\end{figure}

The proton beam line is part of a merged beam instrument dedicated to ion-ion collisions. Fig.~\ref{setup} gives an overview of the optical elements relevant to this work. Multiple apertures (2, 1.2 and 1.1~mm in diameter) are located along the 3.1 m flight path from the exit of the 45 degree deflector to the graphene supporting TEM grid. The latter is in thermal contact with a resistively heated copper block stabilized in temperature. Energy selection is operated by the combination of a Wien filter and a 45-degree cylindrical deflector, which together define the FWHM energy spread to $\Delta E/E \simeq 5 \times 10^{-3}$. The gas cell consists of a 3 cm long cylinder enclosed by two apertures.  Charge exchange is nearly resonant with N$_2$O (IP = 12.89 eV) and CO$_2$ (IP = 13.78 eV), the latter being preferred for safety reasons. The detector is a triple microchannel plate (MCP) stack 40~mm in diameter backed with a resistive anode (Quantar MCP/RAE 3395), coupled with a 10-bit position encoder. Fast timing is ensured by high-pass filtering the high voltage signal and passing it through a combined amplifier and constant fraction discriminator (ORTEC). All timing signals are processed by a 16-channel, 120 ps resolution, time-to-digital module (TDC-V4, DTPI, Orsay). 

\begin{figure}[h!]
	\centering
	\includegraphics[width=0.6\columnwidth]{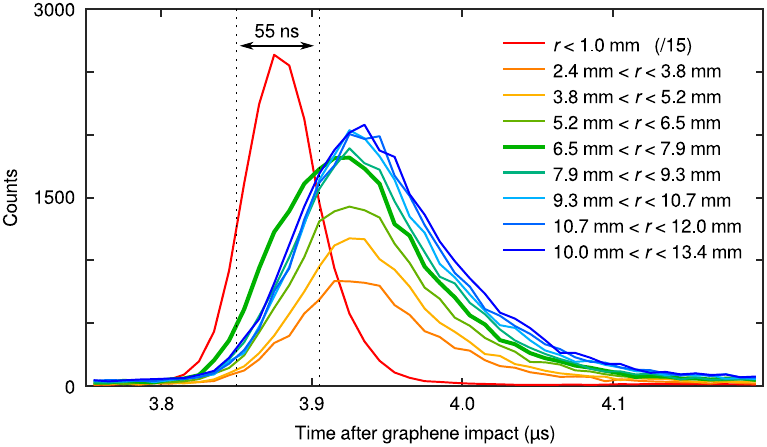}
	\caption{Time-of-flight distribution of events recorded at 300 eV, as a function of radial coordinate in the image centered on the ballistic peak. The thick line corresponds to the location of diffraction events. }
	\label{TOF}
\end{figure}

The time-of-flight (TOF) measurement is performed by recording the time delay between the voltage pulse and the impact of the hydrogen atom on the MCP. 
The beam gating is performed at 800~kHz with 80~ns square pulses, while the time of flight (16~{\textmu}s at 300 eV) is significantly longer than the pulsing period (1.25~{\textmu}s). The time-to-digital converter (TDC) is started by the atom impact and stopped by the leading edge of the next voltage pulse, and that time interval is stored in list mode together with the spatial coordinates of the corresponding impact. The absolute TOF is separately measured at much lower repetition rate with reduced statistics. Due to finite switching time of the voltage applied to the deflector, and the proton beam propagation through the parallel plate condenser it consists of, the actual duration of the beam pulse is a convolution of the beam energy distribution with some apparatus function (see Fig. \ref{TOF}). Subtracting the weighted mean of slices corresponding to radii $5.2<r\leq 6.5$~mm and $7.9<r\leq 9.3$~mm to the slice at $6.5<r\leq 7.9$~mm isolates the contribution of diffraction events, which appear in excess under the ballistic peak. 

\newpage
\section{Suspended graphene on metal grids}

\begin{figure}[!h]
	\centering
	\includegraphics[width=0.7\columnwidth]{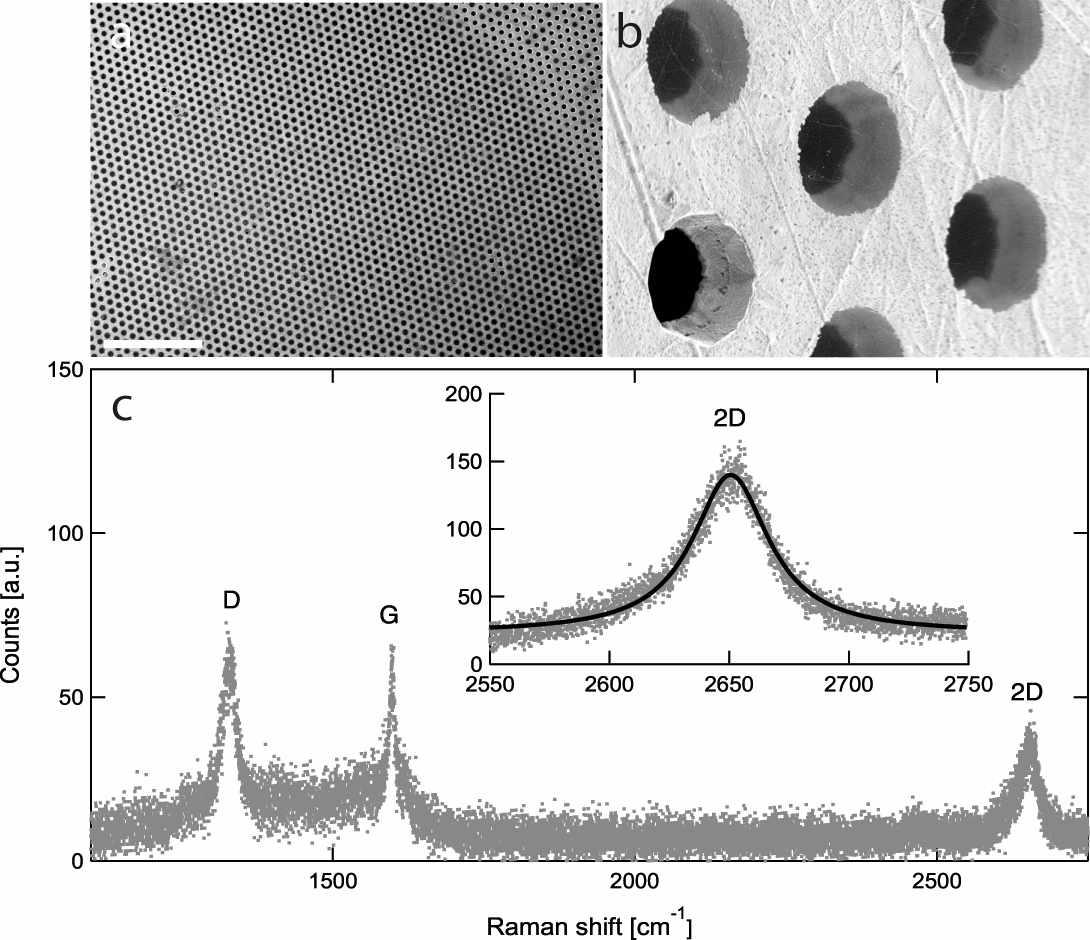}
	\caption{(a) Electron micrograph of part of one of the graphene-covered TEM grids. The scale bar (white) is 100~{\textmu}m. (b) Electron micrograph (tilted view) of several graphene-covered holes in the TEM grid. Each hole has a 6.5~{\textmu}m-diameter. The lower left hole in the micrograph is not covered with graphene, contrary to the other holes. (c) Typical Raman spectrum acquired on a graphene-covered hole (see the text for the parameters), after the diffraction experiments. The typical graphene bands (D, G and 2D) are indicated. Inset: high-resolution spectrum limited to the region of the 2D peak. The continuous black line is a Lorentzian fit to the 2D peak (FWHM: $19.4 \pm 2$ cm$^{-1}$). }
	\label{SEM_raman}
\end{figure}

The samples used in the experiments described in the main text were acquired from Ted Pella and are described as follows: PELCO\textregistered\ Single Layer Graphene on Ultra-fine 2000 Mesh Copper TEM Grid supported by $1\times 2$~mm Synaptek\textsuperscript{TM} Slotted Grid. These samples consist of single-layer polycrystalline graphene grown using a chemical vapor deposition process and transferred using a wet process on top of a copper grid with a regular array of 6.5~{\textmu}m-diameter holes. While the coverage of graphene on the metal grid is supposed to be homogeneous, Fig. \ref{SEM_raman}(a)  shows that a fraction of holes in the grid are not covered with graphene (e.g. top right part of the electron micrograph, with a lighter gray contrast). A close-up (tilted) electron micrograph of the grid, shown in Fig. \ref{SEM_raman}(b), reveals that graphene covering the holes is decorated with small metallic particles. In some graphene regions, the particles form intersecting line patterns, most probably revealing the position of lines of defects or grain boundaries in graphene. 

{\it Raman spectroscopy} -- Raman micro-spectroscopy was performed using a commercial LabRam HR confocal system from Horiba, equipped with a red laser (633 nm) excitation source and a 100× objective (numerical aperture: 0.95). We ensured that the laser power reaching the sample always remained below 1 mW. For broad spectral range measurements (Fig. \ref{SEM_raman}(c), a grating of 150 lines/mm was used, while for more precise measurements focusing on a smaller range, a grating of 2400 lines/mm was employed. It was not possible to precisely determine the degree of laser focusing on the surface of suspended graphene: unlike graphene deposited on a Si substrate, where a well-defined laser spot is visible, the laser light here appears diffused, primarily by the small metallic particles. This suggests that the probed area is likely on the order of 1-5 µm in diameter when targeting the center of a hole in the grid. Since the graphene is suspended, with a significant gap between it and the surface beneath the grid -- which reflects part of the laser light (as the grid is deposited on a Si wafer)-- the Raman signal is relatively weak. This explains the relatively high noise level in the data shown in Fig. \ref{SEM_raman}(c).

The Raman signatures of vibrational modes characteristic of graphene are visible in Fig.~\ref{SEM_raman}(c), which shows a typical spectrum measured when centering the laser on one of the graphene-covered holes. The spectrum features a disorder-induced D band centered at $\sim1330$~cm$^{-1}$, a G band centered at 1600~cm$^{-1}$, and a 2D band centered at 2650~cm$^{-1}$. The I(D)/I(G) ratio ranges between 0.5 and 3, depending on the probed region, indicating a relatively high density of defects. As illustrated in the inset of Fig.~\ref{SEM_raman}(c), the 2D peak can be fitted using a single Lorentzian function, with a full width at half maximum (FWHM) typically ranging between 19 and 30~cm$^{-1}$, depending on the position in the grid—an indication of monolayer graphene. There is no significant qualitative or quantitative difference in the Raman spectra measured before and after the diffraction experiments. In other words, either no defects were created during the interaction of hydrogen atoms with the suspended graphene or, if defects were generated, their Raman signature remains below the detection limit of our measurements.

{\it Thermal desorption and surface roughness} -- The reappearance of an adsorbed layer when cooling the sample under high vacuum conditions does not support the current understanding of surface contamination of CVD graphene transferred on TEM grids. PMMA is best removed by thermal treatment and is usually thought to trap water molecules. Should this polymer get removed at some point, its recapture by the graphene layer is unlikely. An alternative explanation for the loss of contrast upon cooling could reside in the mechanical properties of graphene deposited on a copper substrate. Indeed, the differing thermal expansion coefficient of graphene (thermal expansion coefficient $\alpha<0$ for T$<350$~K \cite{Tian2016}) and Cu ($\alpha\simeq 16\times10^{-6}$~K$^{-1}$) leads to flattening of the graphene ripples \cite{Meyer2007}  likely exacerbated by the presence of PMMA islands. On the other hand, thermal annealing at high temperature may lead to the formation of ripples upon cooling \cite{Bao2009}.

\section{Diffraction images and energy loss versus kinetic energy}

\begin{table}[!ht]
	\begin{tabular*}{0.7\columnwidth}{c @{\extracolsep{\fill}} c @{\extracolsep{\fill}} c @{\extracolsep{\fill}} c}  
		\hline\hline
		\multicolumn{4}{c}{H$^0$ energy (eV)}\\
		Nominal        & Ballistic   & Inelastic   & Elastic \\
		\hline
		300       & 299.96(1)  & 290.57(2)   & 299.69(4) \\
		600       & 592.54(5)  & 578.63(10)  & 591.78(42)  \\
		1200      & 1214.73(10)& 1194.80(18) & 1214.11(1.02) \\
		\hline\hline
	\end{tabular*}
	
	\caption{Average kinetic energy of H atoms after transmission through the graphene sheet, as obtained from the correlation of TOF with position of impact. Numbers in parentheses represent 1$\sigma$ uncertainty of the mean in units of the last significant digits.}
	
	\label{tbl:energy-loss}
	
\end{table}
\begin{table}[!ht]
	\begin{tabular*}{0.5\columnwidth}{c @{\extracolsep{\fill}} c @{\extracolsep{\fill}} c @{\extracolsep{\fill}} c}  
		\hline\hline
		\multicolumn{3}{c}{H$^0$ energy loss (eV)}\\
		Energy   & Inelastic   & Elastic \\
		\hline
		300       & 9.4(1)  & 0.27(4) \\
		600       & 13.9(1)  & 0.8(4) \\
		1200      & 19.9(2) & 0.6(10) \\
		\hline\hline
	\end{tabular*}
	
	\caption{Average kinetic energy loss of H atoms after transmission through the graphene sheet, as obtained from the correlation of TOF with position of impact. Numbers in parentheses represent 1$\sigma$ uncertainty of the mean in units of the last significant digits.}
	
	\label{tbl:energy-loss2}
	
\end{table}
 \begin{figure}[!h]
 	\centering
 	\includegraphics[width=0.6\columnwidth]{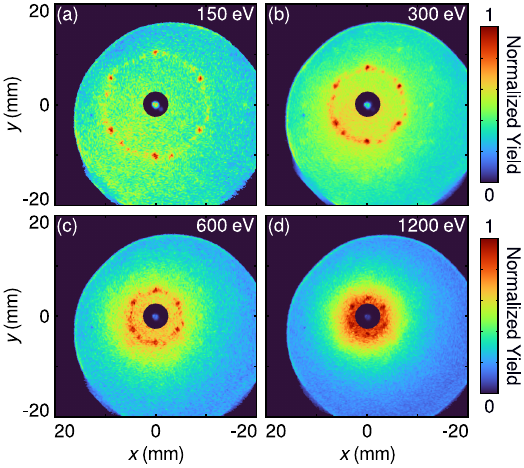}
 	\caption{Diffraction images recorded with the same Ted Pella graphene sample at (a) 150~eV, (b) 300~eV, (c) 600~eV, and (d) 1200~eV. Low energy images are affected by the dark count rate ($\sim 5$~Hz) of comparable magnitude to the signal. In each image, the yield has been normalized to the maximum of the first order diffraction peaks. Counts in the central disk have been divided by 1000 to reveal the spatial distribution of  ballistic events.}
 	\label{energy-scan}
 \end{figure}
\newpage

\section{Debye-Waller factor}
Based on a dark-field transmission electron microscopy study of the Debye-Waller factor (DWF) of graphene \cite{Shevitski2013}, we evaluate the effect of atomic thermal motion on diffraction order attenuation. The Debye-Waller factor (DWF), given by $\exp(-2W)$, accounts for lattice disorder due to temperature. Under certain approximations, detailed in \cite{Shevitski2013}, $2W$ becomes
\begin{align}
	2W\approx G^2 \left[ \frac{\hbar}{k_DM v_s}+ \frac{2k_B T}{k_D^2 M v_s}\log\left(\frac{k_B T}{\hbar v_s k_s} \right) \right]\;,
	\label{eq:dwf}
\end{align}
where $k_\mathrm{B}$ is the Boltzmann constant, $T$ is the temperature, $M$ is the mass of a carbon atom, $v_s$ is the in-plane sound velocity of graphene, $k_\mathrm{D}$ is the Debye wave vector, $k_s=2\pi/L$ is the wave vector, where $L$ is the size of the crystal and $G$ is the reciprocal lattice vector. The quantity between square brackets is the mean-square in-plane displacement, $u_p^2$, which amounts to 44~pm$^2$ at 300~K for a crystal size $L=10~\mu$m.

\begin{figure}[h!]
	\centering
	\includegraphics[width=0.6\columnwidth]{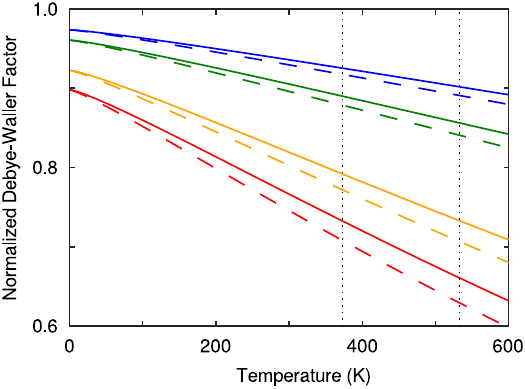} 

	\caption{Debye-Waller factor (DWF) relative to the first diffraction order, as a function of temperature. Full lines: $L=2\times 10^{-6}$~m ({\it Graphenea}); dashed lines: $L=6.5\times 10^{-6}$~m ({\it Ted Pella}).  $G/G_{min}=\sqrt{3}$ (blue), 2 (green), $\sqrt{7}$ (orange), and 3 (red).}\label{fig:dwf}    
\end{figure}

The DWF has been computed using the following values: $M = 1.99\times 10^{-26}$~kg, $v_s=2.20\times 10^4$~m/s, $k_\mathrm{D}=1.55 \times 10^{10}$~m$^{-1}$ and $k_s=6.28 \times 10^{5}$~m$^{-1}$ with $L= 2\times 10^{-6}$~m and $6.5 \times 10^{-6}$~m, corresponding to the hole diameter of {\it Graphenea} and {\it Ted Pella} supporting grids, respectively. If we normalize the DWF to the first diffraction peak \textit{i.e.} $G=1$ in units of $G_{min}=4\pi/\sqrt{3}a$, with $a=142$~pm the graphene lattice parameter, we get the attenuation of the other peaks with respect to the first one. As shown in Fig. \ref{fig:dwf} (dashed lines), the normalized DWF rapidly decreases with temperature, the more so for higher diffraction orders. 

\begin{table}[h!]
	\begin{tabular*}{0.7\columnwidth}{l @{\extracolsep{\fill}} c @{\extracolsep{\fill}} c @{\extracolsep{\fill}} c @{\extracolsep{\fill}} c @{\extracolsep{\fill}} c}
		\hline\hline
		\multicolumn{6}{c}{$300~$eV, {\it Graphenea},  $100~^\circ$C}\\
		$G/G_{min}$        & Experiment & normalized DWF & {\it ab initio} & H-C binary & Brand \etal \cite{Brand2019}\\
		\hline
		$1$        & 1 & 1 & 1 & 1 & 1 \\
		$\sqrt{3}$ & 0.241(32) & 0.925 & 0.147 & 0.421 & 0.489 \\
		$2$        & 0.172(32) & 0.890 & 0.178 & 0.755 & 0.553 \\
		\hline\hline
		\multicolumn{6}{c}{$600~$eV, {\it Graphenea}, $100~^\circ$C}\\
		$G/G_{min}$        & Experiment & normalized DWF & {\it ab initio} & H-C binary & Brand \etal \cite{Brand2019}\\
		\hline
		$1$        & 1 & 1 & 1 & 1 & 1 \\
		$\sqrt{3}$ & 0.258(32) & 0.925 & 0.154 & 0.381 & 0.496 \\
		$2$        & 0.168(29) & 0.890 & 0.138 & 0.376 & 0.229 \\
		$\sqrt{7}$ & 0.164(26) & 0.792 & 0.075 & 0.118 & 0.122 \\
		$3$        & 0.114(27) & 0.732 & 0.136 & 0.212 & 0.185 \\

		\hline\hline
	\end{tabular*}
	\caption{Relative intensity of the lowest diffraction orders. Theoretical values have been multiplied by the Debye-Waller factor computed with $L$ corresponding to the sample type and $T$ the temperature at which the corresponding images were recorded. Numbers in parentheses represent 1$\sigma$ uncertainty of the mean in units of the last significant digits.}
	\label{tbl:theo-exp-diffraction-order-debye-waller}
\end{table}

\section{{\sc Quantum Espresso} parameters}
The convergence threshold of the self-consistent field method was set at $10^{-9}$ Ry, ensuring that the self-consistency criterion was met with a very high degree of accuracy, critical for obtaining reliable results. The calculated interaction potential was evaluated on a discrete grid of points. The number of grid points ($N_x=23$, $N_y=90$ and $N_z=11$) as well as the spacing ($\Delta x \approx \Delta y \approx 0.025$~{\AA}, and $\Delta z = 0.2$~{\AA}) were carefully adjusted in order to obtain converged results and a sufficiently fine description of the graphene lattice. The calculations were performed at HPC Center of the University of Strasbourg.

\newpage
	\bibliography{bibliography}